\documentclass[11pt]{article}
\usepackage{times,amssymb,amsmath,amsfonts,graphicx}
\usepackage{amssymb,amsmath,latexsym,graphics,epstopdf}
\usepackage{epsfig,enumerate,amscd}
\usepackage{color,theorem}
\usepackage{bm}

\newtheorem{thm}{Theorem}[section]

\newtheorem{prop}[thm]{Proposition}
\newtheorem{coro}[thm]{Corollary}

\theorembodyfont{\upshape}
\newtheorem{rem}[thm]{Remark}

\newtheorem{assumption}[thm]{Assumption}

\newcommand{\proof}{{\noindent\it Proof.}\quad}
\newcommand{\finproof}{{\hfill$\Box$\\}}
\newcommand{\bb}[1]{{\mathbb #1}}
\newcommand{\E}{{\bb E}}
\newcommand{\esp}{{\bb E}}

\newcommand{\R}{{\mathbb{R}}}
\newcommand{\Y}{{\mathbb{P}}}

\newcommand{\F}{{\mathcal{F}}}

\newcommand{\di}{{\mathrm{d}}}
\newcommand{\G}{{\mathcal{G}}}
\newcommand{\FF}{{\mathbb{F}}}
\newcommand{\bF}{{\mathbb{F}}}

\newcommand{\bG}{{\mathbb{G}}}
\newcommand{\Q}{{\mathbb{Q}}}

\newcommand{\1}{1\hspace{-2.6pt}\mathrm{l}}
\newcommand{\indic}{1\hspace{-2.6pt}\mathrm{l}}

\newcommand{\bmu}{\bm{\mu}}
\newcommand{\bsigma}{\bm{\sigma}}
\newcommand{\bgamma}{\bm{\gamma}}
\usepackage{color}
\definecolor{Ying}{rgb}{0.5,0,1.0}

\definecolor{Yang}{rgb}{1.0,0.0,1.0}

\makeatletter
\@addtoreset{equation}{section}

\makeatother

\begin{document}
\thispagestyle{empty}
\title{Credit derivatives pricing with default density term structure modelled by L\'evy random fields}
\author{Lijun Bo\thanks{Department of Mathematics, Xidian University, Xi'an 710071 China. 
}\quad Ying Jiao\thanks{Laboratoire de Probabilit\' es et Mod\`eles Al\' eatoires, Universit\'e Paris Diderot--Paris 7,  Paris 75013 France, and School of Mathematical Sciences and BICMR, Peking University, Beijing 100871 China. 
}\quad Xuewei Yang\thanks{School of Mathematical Sciences, Nankai University,  94 Weijin Road, Nankai District, Tianjin 300071 China.
       Funding: NSF of China (No. 11101223)}}
\date{\today}
\maketitle

\begin{abstract}
We model the term structure of the forward default intensity and the default density by using L\'evy random fields, which allow us to consider the credit derivatives with an after-default recovery payment. As applications, we study the pricing of a defaultable bond and represent the pricing kernel as the unique solution of a parabolic integro-differential equation. Finally, we illustrate by numerical examples the impact of the contagious jump risks on the defaultable bond price in our model.\\
\vspace{1mm}

\end{abstract}
\section{Introduction }
\label{sec:Intro}

The term structure modelling in the interest rate and in the credit risk modelling has been widely adopted and extended
since the original paper of Heath-Jarrow-Morton \cite{Heath_Jarrow_Morton}.
Notably, there have appeared many important papers (e.g.  \cite{Albeverio_Lytvynov_Mahnig,
Bjork_Kabanov_Runggaldier,
EberleinJacodRaible2005,
EberleinRaible1999,
FilipovicTappe2008,
FilipovicTappeTeichmann2010})
incorporating jump diffusions to describe the family of bond prices
or the forward curves as a generalization of the classic HJM model.

In the credit risk modelling, the conditional survival probability
associated to the default time is an important quantity for
measuring default risk and studying valuation of credit derivatives.
Let  $\tau$ be a nonnegative random variable defined on
a complete probability space $(\Omega,\mathcal{F},\mathbb{P})$
equipped with a filtration $\bF=(\F_t)_{t\geqslant 0}$ satisfying
the usual conditions. 
The  conditional survival probability (CSP)
is defined as $S_t(\theta)=\Y(\tau>\theta|\F_t)$, $t,\theta\geqslant 0$. 
To describe the term structure of the CSP, we can use both the
density and the intensity point of view. On the one hand, as in El
Karoui et al \cite{El_Jeanblanc_Jiao_2010}, we assume that there
exists a family of  $\F_t\otimes{\mathcal{B}}(\R_+)$-measurable
functions $(\omega,\theta)\to \alpha_t(\omega,\theta)$ such that the
CSP has the following additive representation:
\begin{gather}\label{model:csp1}
S_t(\theta)=\int_\theta^\infty\alpha_t(v)\di v.
\end{gather}
The family of random variables $\alpha_t(\cdot)$ is called the
conditional {\it density} of the default time $\tau$ given $\F_t$.  On the other hand,
similarly to the definition of forward rate, we can use the
``intensity'' point of view and the following multiplicative
representation:
\begin{gather}\label{def:forward-intensity}
S_t(\theta)=\exp\left(-\int_0^\theta \lambda_t(v)\di v\right)
\end{gather}
where the $\F_t\otimes{\mathcal{B}}(\R_+)$-measurable function
$(\omega,\theta)\to \lambda_t(\omega,\theta)$ is called the {\it
forward intensity}. It is equivalent to assume the existence of the
density or the intensity for all positive $t$ and $\theta$. We have
the relationship:
\begin{gather}\label{eq:relation}
\alpha_t(\theta)=S_t(\theta)\lambda_t(\theta).
\end{gather}
In the interest rate models, the time $\theta$ is always larger than $t$ and the forward rate
has no economic interpretation for $\theta<t$. However, it is noted
in \cite{El_Jeanblanc_Jiao_2010} that to study what happens after a
default event, we need the whole term structure of the conditional
survival probability, that is, for all positive $t$ and $\theta$.
One typical example is a defaultable bond where the recovery payment
is effectuated at a given maturity later than the economic default
date.

In this paper, we consider the whole term structure modelling of CSP and
the applications to the credit derivative pricing. In the credit
risk models, the default contagion phenomenon is often modelled by
positive jumps in the intensity process. We take this point into
modelling consideration and propose a forward intensity driven by
L\'evy random fields. In the existing L\' evy term structure models,
in Filipovi\'c et al. \cite{FilipovicTappe2008,
FilipovicTappeTeichmann2010}, the authors consider forward
curve evolutions as 
solutions of the infinite dimensional Musiela parametrization
first-order hyperbolic stochastic differential equations driven by
$n$-independent
L\'evy processes 
or driven by a Wiener process together with an
independent Poisson measure. 
In \cite{EberleinRaible1999}, Eberlein and Raible present a class
of bond price models that can be driven by a wide range of
L\'evy processes with finite exponential moments. This
model was further applied to describe the defaultable L\'evy term
structure and explore ratings and restructuring of the defaultable
market. The driving process of the L\'evy term structure model in
\cite{EberleinRaible1999} was further extended to non-homogeneous
L\'evy processes in \cite{EberleinJacodRaible2005}.

Motivated by those existing L\'evy term structure models and the
random field models which are widely used to model various stochastic dynamics
(e.g. \cite{Albeverio_Lytvynov_Mahnig, Dalang,
Dalang_Frangos,
DalangQuer-Sardanyons, Goldstein2000, Kennedy1994, Kennedy1997}),
we suppose that the L\'evy random
field in our model is a combination of a kernel-correlated Gaussian
field and an independent (central) Poisson random measure. The jump
component described as Poisson measure is similar to that used in
\cite{FilipovicTappeTeichmann2010}, but it is not necessary to
assume the exponential integrability condition for the
characteristic measure under our framework (see Section~2). The kernel-correlated Gaussian field is more flexible compared to
the Gaussian components without kernel-correlation considered in
\cite{Goldstein2000,Kennedy1994,Kennedy1997}. In fact, we can choose
appropriate correlated-kernels of the Gaussian field so that the
models considered in
\cite{EberleinRaible1999,FilipovicTappe2008,FilipovicTappeTeichmann2010}
can be covered (see Remark~\ref{rem:comparison}). Note that it is not
genetically tractable for pricing of defaultable bonds under
infinite dimensional framework as in \cite{FilipovicTappe2008,
FilipovicTappeTeichmann2010}. Although we do not intend to consider
the forward intensity under infinite dimensional framework as in
\cite{FilipovicTappe2008, FilipovicTappeTeichmann2010}, it has a
close relationship between the (infinite dimensional) Wiener process
and the kernel-correlated Gaussian field. Indeed, the
kernel-correlated Gaussian field can product a cylindrical Wiener
process  by establishing appropriate Hilbert spaces (see Proposition
2.5 in \cite{DalangQuer-Sardanyons}). We deduce the dynamics of the
CSP and the associated density in this setting. In particular, we
emphasize on a martingale condition, which can be viewed as an
analogue of the non-arbitrage condition in the classical HJM model.

For the pricing of credit derivatives, we follow the standard general framework in Bielecki and Rutkowski \cite{Bielecki_Rutkowski_2002}.
The global market information
contains both the default information and the ``default-free''
market information represented by  the filtration $\bF$, which is obtained by an
enlargement of filtration. We are particularly interested in an
economic default case, that is, the default does not lead to the
total bankruptcy of the underlying firm and a partial recovery value
is repaid at the maturity date of the bond in case of default prior to the maturity. To
evaluate this ``after-default'' payment, we use the density approach
in \cite{El_Jeanblanc_Jiao_2010} and obtain that the key quantities for the pricing of a defaultable bond are two
pricing kernels, one depending on the interest rate and the default
density, and the other depending on additionally the recovery
rate. We assume that both the short interest rate and the default
density are modelled by the L\' evy random field model and are
correlated between them.  For the recovery rate, we analyze firstly
the simple case where the recovery rate is deterministic and then
the random recovery case. We show that the pricing kernel is related
to the solution of a second-order parabolic integro-differential
equation and we prove, based on a result of Garroni and Menaldi
\cite{Garroni_Menaldi}, the existence and the uniqueness of the
solution to the equation.

The rest of the paper is organized as follows. We present our model setting in Section \ref{forward intensity} and
give the martingale condition. We then analyze the dynamics of the CSP and the conditional density in Section \ref{CSP}.
In Section \ref{Sec: pricing}, we discuss the pricing of credit derivatives and in particular the defaultable zero-coupon
bond. The two sections \ref{sec:pring-kernel1} and \ref{sec:pring-kernel2} focus on the pricing kernels. Finally, we present
some numerical illustrations in the last section \ref{sec:numerical}.


\section{Forward intensity driven by L\'evy random field}
\label{forward intensity}

In this paper, we adopt a random field point of view to model
the forward intensity $\lambda_t(\theta)$ where both $t$ and $\theta$ are positive. We consider a L\'evy random field on $\mathbb R_+\times\mathbb R^d$ which is a combination of a Gaussian random field $Y^G$ and a compensated Poisson random measure $Y^P$ independent to $Y^G$. Here $\mathbb R_+$ denotes the time space and $\mathbb R^d$ is considered as a parameter space.

We assume that the covariance of the Gaussian random field $Y^G$ is given by a kernel measure $c$ on $\mathbb R^d$ which has a continuous and symmetric density on $\mathbb R^d\setminus\{0\}$ with respect to the Lebesgue measure and such that $c(\{0\})>0$. Namely for $(\phi_1,\phi_2)\in C_0^{\infty}(\mathbb R_+\times\mathbb R^d)^2$,
\[\mathbb E[Y^G(\phi_1)Y^G(\phi_2)]=\int_{0}^\infty\int_{\mathbb{R}^{2d}}\phi_1(t,\xi_1)
\phi_2(t,\xi_2)c(\xi_1-\xi_2)\mathrm{d}\xi_1\mathrm{d}\xi_2\mathrm{d} t,\]
where by abuse of language $c(\xi_1-\xi_2)\mathrm{d}\xi_1\mathrm{d}\xi_2$ denotes the measure on $\mathbb R^d\times\mathbb R^d$ the inverse image of the measure $c$ by the mapping from $\mathbb R^d\times\mathbb R^d$ to $\mathbb R^d$ which sends $(\xi_1,\xi_2)$ to $\xi_1-\xi_2$.
The Gaussian random field $Y^G$ defines a worthy martingale measure (see \cite[p.289]{Walsh1986} and \cite[p.190]{Dalang_Frangos}). 
Let $\mathbb F^G=(\mathcal F_t^G)_{t\geqslant 0}$ be the filtration
satisfying the usual conditions which is generated by
\[\sigma(Y^G([0,u]\times A),\;u\leqslant t,\; A\in\mathcal
B_b(\mathbb R^d)),\quad t\geqslant 0,\] where $\mathcal B_b(\mathbb
R^d)$ denotes the set of all bounded Borel subsets of $\mathbb R^d$.
Let $\mathcal P^G$ be the predictable $\sigma$-algebra on
$\Omega\times\mathbb R_+$ associated to $\mathbb F^G$
and $\Phi_c$ be the linear space of all $\mathcal P^{G}\otimes\mathcal B(\mathbb R^d)$-measurable functions $h$ such that
\[\|h\|_{c,T}:=\mathbb E\bigg[\int_0^T\int_{\mathbb R^{2d}}|h(t,\xi_1)h(t,\xi_2)|c(\xi_1-\xi_2)\mathrm{d}\xi_1\mathrm{d}
\xi_2\mathrm{d}t\bigg]^{1/2}<+\infty\] for any $T>0$. The stochastic
integral $h\cdot Y^G$ is well defined for any $h\in\Phi_c$. When $c$
is the Dirac distribution concentrated on the origin, the stochastic
integral:
\begin{gather}\label{def:Brownian-sheet}
B(t_0,\ldots,t_d)=Y^G([0,t_0]\times\cdots\times[0,t_d]),\quad
(t_0,\ldots,t_d)\in\mathbb R_+^{d+1}
\end{gather}
defines a $(d+1)$-parameter Brownian sheet. If in particular $d=0$,
it becomes a standard Brownian motion.


Denote the intensity measure of the compensated Poisson field $Y^P$
by $\nu(\mathrm{d}\xi)\mathrm{d}t$, $(t,\xi)\in\mathbb
R_+\times\mathbb R^d$, where $\nu$ is a $\sigma$-finite measure on
$\mathbb R^d$. Let $\mathbb F^P=(\mathcal F^P_t)_{t\geqslant 0}$ be
the filtration satisfying the usual conditions generated by
\[\sigma(Y^P([0,u]\times A),\;u\leqslant t,\;A\in\mathcal B_b(\mathbb R^d)),\quad t\geqslant 0\]
and $\mathcal P^P$ be the predictable $\sigma$-algebra on $\Omega\times\mathbb R_+$ associated to $\mathbb F^P$.
Denote by $\Psi_\nu$ the linear space of all $\mathcal P^P\otimes\mathcal B(\mathbb R^d)$-measurable functions such that
\[\|g\|_{\nu,T}:=\mathbb E\bigg[\int_0^T\int_{\mathbb R^d}|g(t,\xi)|^2\nu(\mathrm{d}\xi)\mathrm{d}t\bigg]<+\infty\]
for any $T>0$. The stochastic integral $g\cdot Y^P$ is well defined for any $g\in\Psi_\nu$.

Let $\mathbb F=(\mathcal F_t)_{t\geqslant 0}$ be the natural
filtration generated by the L\'evy random field, namely $\mathbb
F:=\mathbb F^G\vee\mathbb F^P$. We describe the forward intensity by
using the L\'evy random filed as  the following additive HJM type
model:
\begin{eqnarray}\label{model:levy-forward-intensity0}
\di \lambda_t(\theta)=\mu_t(\theta)\di t + \int_{\R^d}\sigma_t(\theta,\xi)Y^{G}(\di t,\di\xi)
+ \int_{\R^d}\gamma_{t-}(\theta,\xi)Y^{P}(\di t,\di\xi),
\end{eqnarray}
where
\begin{enumerate}[(1)]
\item ${{\bmu}}=(\mu_t(\theta);\ (t,\theta)\in\R_+^2)$ is $\mathcal P\otimes\mathcal B(\mathbb R_+)$-measurable and $\int_0^T {\E}\left[|\mu_t(\theta)|\right]\di t<\infty$, where $\mathcal P$ is the predictable $\sigma$-algebra on $\Omega\times\mathbb R_+$ associated to the filtration $\mathbb F$,
\item ${{\bsigma}}=(\sigma_t(\theta,\xi);\
(t,\theta,\xi)\in\R_+\times\mathbb R_+\times\R^d)$ is $\mathcal P^G\otimes\mathcal B(\mathbb R_+\times\mathbb R^d)$-measurable and for any $\theta\in\mathbb R_+$, $\sigma_{\cdot}(\theta,\cdot)\in\Phi_c$,
\item ${\bgamma}=(\gamma_t(\theta,\zeta){\geqslant 0};\
(t,\theta,\zeta)\in\R_+\times \R_+\times\R^d)$ is $\mathcal P^P\otimes\mathcal B(\mathbb R_+\times\mathbb R^d)$-measurable and for any $\theta\in\mathbb R_+$, $\gamma_\cdot(\theta,\cdot)\in\Psi_\nu$.
\end{enumerate}
The model \eqref{model:levy-forward-intensity0} can also be written in the integral form as
\begin{eqnarray}\label{model:levy-forward-intensity1}
\lambda_t(\theta)=\lambda_0(\theta)+\int_0^t\mu_s(\theta)\di s +
\int_0^t\int_{\R^d}\sigma_s(\theta,\xi)Y^{G}(\di s,\di\xi)+
\int_0^t\int_{\R^d}\gamma_{s-}(\theta,\xi)Y^{P}(\di s,\di\xi)
\end{eqnarray}
where both stochastic integrals with respect to $Y^G$ and $Y^P$ are $\mathbb F$-martingales with mean zero, $\lambda_0(\cdot)$ is a deterministic Borel function on $\mathbb R_+$.

Similarly to the classical HJM model, in the above L\'evy field
model \eqref{model:levy-forward-intensity0}, there exists a
relationship between the drift coefficient $\bmu$ and the diffusion
coefficients $\bsigma$ and $\bgamma$ due to the fact that, for any
$\theta\geqslant 0$, the conditional survival probability process
$\big(S_t(\theta)=\exp\big(-\int_0^\theta\lambda_t(v)\di
v\big),\;t\geqslant 0\big)$ should be an $\mathbb F$-martingale. We
call this relationship the martingale condition ({\bf MC}). Let us
introduce the following notation:
\begin{eqnarray*}
I_\mu(t,\theta):=\int_0^\theta\mu_t(v)\di v,\ \ \ I_\sigma(t,\theta,\xi):=\int_0^\theta\sigma_t(v,\xi)\di v,
\ \ \ {\rm and}\ \ I_\gamma(t,\theta,\xi):=\int_0^\theta\gamma_t(v,\xi)\di v,
\end{eqnarray*}
where $(t,\theta,\xi)\in\R_+\times\mathbb R_+\times\R^d$.

\begin{thm}
For all $\theta\geqslant 0$ and $T>0$, one has
\begin{eqnarray}\label{eq:four statements}
\int_0^T {\E}\left[|I_\mu(t,\theta)|\right]\di t<\infty,\ \
I_\sigma(\cdot,\theta,\cdot)\in\Phi_c,\ \ I_\gamma
(\cdot,\theta,\cdot)\in\Psi_\nu,\ \ {\rm and}\ \
e^{-I_{\gamma}({\cdot},\theta,\cdot)}-1\in \Psi_\nu,
\end{eqnarray}
Moreover,  the process family $\big(S_t(\theta)=\exp\big(-\int_0^\theta\lambda_t(v)\di v\big),\;t\geqslant 0\big)$ is a family of  $\mathbb F$-martingales if and only if the following condition is satisfied:
\begin{equation}\label{eq:S-martingale-condition}
\begin{split}
{\bf(MC)}\qquad \forall\ \theta\geqslant 0,\,\, \, \mu_t(\theta)
=&\int_{\R^{2d}}\sigma_t(\theta,\xi_1)I_{\sigma}(t,\theta,\xi_2)c(\xi_1-\xi_2)\di\xi_1\di\xi_2\\
&+\int_{\R^d}\gamma_t(\theta,\xi)(1-e^{-I_{\gamma}(t,\theta,\xi)})\nu(\di\xi).
\end{split}\end{equation}

\end{thm}
\proof   The proofs for the first three statements in \eqref{eq:four statements} are similar. We only provide the details for the third one. For any $T>0$, we have
\[\begin{split}
\int_0^T\int_{\mathbb{R}^{d}} {\bb E}|I_\gamma
(t,\theta,\xi)|^2\nu(\di\xi)\di t
&= \int_0^T\int_{\mathbb{R}^{d}} {\bb E}\left|\int_0^\theta\gamma_t(v,\xi)\di v\right|^2\nu(\di\xi)\di t\\
&= \int_0^T\int_{\mathbb{R}^{d}} {\bb E}\left[\int_0^\theta\gamma_t(v_1,\xi)\di v_1\int_0^\theta\gamma_t(v_2,\xi)\di v_2\right]\nu(\di\xi)\di t\\
&\leq  \frac{1}{2}\int_0^T\int_0^\theta\int_0^\theta\int_{\mathbb{R}^{d}} {\bb E}\left[|\gamma_t(v_1,\xi)|^2+|\gamma_t(v_2,\xi)|^2\right]\nu(\di\xi)\di v_1\di v_2\di t\\
&=  \theta \int_0^\theta \int_0^T \int_{\mathbb{R}^{d}} {\bb
E}\left[|\gamma_t(v,\xi)|^2\right]\nu(\di\xi)\di t\di v,
\end{split}\]
which is finite since $\gamma_{\cdot}(v,\cdot)\in\Psi_{\nu}$ for any
$v\geqslant 0$. For the last assertion in \eqref{eq:four
statements}, note that $\gamma_t(\theta,\xi)\geqslant 0$ and thus
$$\left|e^{-I_{\gamma}({t},\theta,\xi)}-1\right|\leq
\left|I_{\gamma}({t},\theta,\xi)\right|,
$$
for all $(t,\theta,\xi)\in\R_+\times\mathbb R_+\times\R^d$.

We now prove that the condition $({\bf MC})$ is equivalent to the martingale condition for $(S_t(\theta),\;t\geqslant 0)$.
In fact\begin{equation}\label{eq:csp-ito}
\begin{split}
\frac{\di S_t(\theta)}{S_{t-}(\theta)}&=-I_\mu(t,\theta)\di t-\int_{\R^d}I_{\sigma}(t,\theta,\xi)Y^G(\di t,\di\xi)\\
&\quad+\frac{1}{2}\int_{\R^{2d}}I_\sigma(t,\theta,\xi_1)I_{\sigma}(t,\theta,\xi_2)c(\xi_1-\xi_2)\di\xi_1\di\xi_2\di t\\
&\quad+\int_{\R^d}(e^{-I_{\gamma}({t^-},\theta,\xi)}-1)Y^P(\di t,\di\xi)\\
&\quad
+\int_{\R^d}(e^{-I_{\gamma}(t,\theta,\xi)}-1+I_\gamma(t,\theta,\xi))\nu(\di\xi)\di
t,
\end{split}\end{equation}
so the martingale condition of $(S_t(\theta),\;t\geqslant 0)$ is thus equivalent to the following equality
\[\begin{split}
I_\mu(t,\theta)
&=\frac{1}{2}\int_{\R^{2d}}I_\sigma(t,\theta,\xi_1)I_{\sigma}(t,\theta,\xi_2)c(\xi_1-\xi_2)\di\xi_1\di\xi_2\\
&\quad+\int_{\R^d}(e^{-I_{\gamma}(t,\theta,\xi)}-1+I_\gamma(t,\theta,\xi))\nu(\di\xi),
\end{split}\]
which is equivalent to $({\bf MC})$.\hfill$\Box$\\

\begin{rem}
Consider the particular case where $d=0$, $c$ is the Dirac measure, and $\nu=0$. The condition ({\bf MC}) becomes
\[\forall\,\theta\geqslant 0,\quad
\mu_t(\theta)=\sigma_t(\theta)\int_0^\theta\sigma_t(v)\di v.\]
 This corresponds to the non-arbitrage condition in the classical HJM model where the forward intensity is driven by a standard Brownian motion.
\end{rem}

\begin{rem}\label{rem:comparison}

There exist random field models in the literature. We make below
some comparisons. The forward intensity model (2.2) can be extended
to the following form:
\begin{equation}\label{model:extended}\begin{split}
\di\lambda_t(\theta)=\mu_t(\theta)&\di t + \int_{\R^d}\sigma_t(\theta,\zeta)
Y^G(\di t,\di\zeta) +
\int_{0<|\xi|\leqslant 1}\gamma_{t-}(\theta,\xi)Y^P(\di t,\di\xi)\\&
+\int_{|\xi|> 1}\hat{\gamma}_{t-}(\theta,\xi)(Y^P(\di t,\di\xi)+\nu(\di\xi)\di t),
\end{split}
\end{equation}
where $\sigma_\cdot(\theta,\cdot)\in\Phi_c$, $\gamma_\cdot(\theta,\cdot)\1_{\{|\cdot|\leqslant1\}}$ and
$\hat{\gamma}_\cdot(\theta,\cdot)\1_{\{|\cdot|>1\}}\in\Psi_\nu$, for each fixed
$\theta\geqslant0$. Under the model \eqref{model:extended}, the corresponding
martingale condition $({\bf MC})$ will be changed accordingly. We next
consider a special form of the predictable random filed with separable variables:
\[\sigma_t(\theta,\zeta)=\widetilde{\sigma}_t(\theta)\widetilde{\phi}(\zeta),\ \
\gamma_t(\theta,\xi)=\hat{\gamma}_t(\theta,\xi)=\langle\widetilde{\gamma}_t(\theta),\xi\rangle,\ \ \zeta\in\R^d,\ \xi\in\R_+^d
\]
where $(\widetilde{\sigma}_t(\theta);\ (t,\theta)\in\R_+^2)$ is a real-valued predictable
random field, $(\widetilde{\gamma}_t(\theta)=(\widetilde{\gamma}_t^1(\theta),\dots,\widetilde{\gamma}_t^d(\theta));\ (t,\theta)\in\R_+^2)$
is a $\R_+^d$-valued predictable field and $\widetilde{\phi}(\zeta)$ is a deterministic
measurable function on $\R^d$. In this case, the extended model \eqref{model:extended} can be rewritten as
\begin{gather}\label{model:extended2}
\di\lambda_t(\theta)=(\mu_t(\theta)-a)\di t + \sigma_t(\theta)Y^G(\di t,\widetilde{\phi}(\star)) +
\langle\widetilde{\gamma}_t(\theta),\di L_t\rangle,
\end{gather}
where $a\in\R$, $\langle\cdot,\cdot\rangle$ denotes the inner-product on $\R^d$ and
\[\di L_t=a\di t+\int_{0<|\xi|\leqslant 1}\xi Y^P(\di t,\di\xi)+\int_{|\xi|> 1}\xi(Y^P(\di t,\di\xi)+\nu(\di\xi)\di t)
\]
is a non-Gaussian L\'evy process if the characteristic measure $\nu$
is a L\'evy measure. If $\widetilde{\phi}\equiv1$, then
$Y^G(\1_{[0,t]}\times\widetilde{\phi}(\star))$ becomes a Brownian
motion when the correlated-kernel is Dirac. Choose appropriate
smooth function $\widetilde{\phi}$ as in the proof of Proposition
2.5 in \cite{DalangQuer-Sardanyons}, then
$Y^G(\1_{[0,t]}\times\widetilde{\phi}(\star))$ becomes a cylindrical
Wiener process. Thus we recover the L\'evy interest rate term
structure models considered in
\cite{EberleinRaible1999,FilipovicTappe2008,FilipovicTappeTeichmann2010},
if the L\'evy measure $\nu$ satisfies the exponential integrability
condition. We next give a comparison of our L\'evy random field
$Y^G+Y^P$ introduced previously in this section with existing L\'evy fields in literature.

\begin{enumerate}
\item As in \eqref{def:Brownian-sheet}, the field $Y^G+Y^P$ can be reduced to a
Brownian sheet in Walsh \cite{Walsh1986}, when the kernel $c$ is
Dirac and the characteristic measure $\nu=0$ (hence $Y^P=0$);
\item the field $Y^G+Y^P$ becomes a so-called ``colored'' space-time white noise model established by
\cite{Dalang}, when the kernel $c(\xi)=|\xi|^{-\alpha}$ with
$0<\alpha<d$ and $\nu=0$;
\item the fractional
space-time white noise (fractional in space and time in white) used
in \cite{Nualart_Ouknine} corresponds to the field $Y^G+Y^P$ with
the kernel $c(\xi)=h(2h-1)|\xi|^{2h-2}$ with $\frac{1}{2}<h<1$,
$d=1$ and $\nu=0$;
\item the Poisson sheet in \cite{Albeverio_Lytvynov_Mahnig}
corresponds to the field $Y^G+Y^P+\nu(\di\xi)\di t$ with $c=0$ and
$\nu(\xi)=z\delta_1(\di\xi)$ where $z>0$ is single point and
$\delta_1$ is the Dirac measure concentrated at $1$. The Gamma sheet
in \cite{Albeverio_Lytvynov_Mahnig} is the field
$Y^G+Y^P+\nu(\di\xi)\di t$ with $c=0$ and
$\nu(\di\xi)=\frac{e^{-\xi}}{\xi}\1_{\{\xi>0\}}z\di\xi$ where $d=1$
and $z>0$ is a single point.
\end{enumerate}

\end{rem}

\section{Conditional survival probability and density}
\label{CSP} In this section, we concentrate on the conditional
survival probability $(S_t(\theta),\;t\geqslant 0)$ and the
conditional density $(\alpha_t(\theta),\;t\geqslant 0)$. Here we
specify a c\`adl\`ag version of the martingale
$(S_t(\theta),\;t\geqslant 0)$ for any $\theta\geqslant 0$. In fact,
to show that the integral $\int_0^\theta\lambda_t(v)\di v$ defines a
c\`adl\`ag process, we need a stronger assumption on the process
$\lambda(\theta)$ in order to apply Lebesgue's theorem. The
c\`adl\`ag version of $S(\theta)$, if well defined, should have a
universal version of its predictable projection as follows:
\[S_{t-}(\theta)=S_t^{(p)}(\theta)=\exp\left(-\int_0^\theta\lambda_{t-}(v)
\di v\right).\]
Thus
\begin{equation}\label{Equ:Sttheta}S_t(\theta):=\lim_{\begin{subarray}{c}
q\in\mathbb Q^+\\
q\downarrow t
\end{subarray}}\exp\left(-\int_0^\theta\lambda_{q-}(v)
\di v\right)\end{equation}
defines a universal c\`adl\`ag version of the martingale $S(\theta)$.

We observe from the equality \eqref{eq:csp-ito} that, under the condition $({\bf MC})$, the conditional survival probability
admits the following dynamics:
\begin{eqnarray}\label{dynamics:csp}
\frac{\di S_t(\theta)}{S_{t-}(\theta)} =
-\int_{\R^d}I_{\sigma}(t,\theta,\xi)Y^G(\di t,\di\xi)+\int_{\R^d}(e^{-I_{\gamma}({t^-},\theta,\xi)}-1)Y^P(\di
t,\di\xi),
\end{eqnarray}
where $S_0(\theta)=\exp(-\int_0^\theta\lambda_0(v)\di v)$.

For $\theta\geqslant 0$, we denote by $M(\theta)$ the martingale defined as
\begin{equation}\label{M t theta}\di M_t(\theta)=-\int_{\R^d}I_{\sigma}(t,\theta,\xi)Y^G(\di t,\di\xi)+\int_{\R^d}(e^{-I_{\gamma}({t-},\theta,\xi)}-1)Y^P(\di
t,\di\xi),\quad M_0(\theta)=0.\end{equation} With this notation,
$S(\theta)/S_0(\theta)$ is the Dol\'eans-Dade exponential of the
martingale $M(\theta)$. Moreover, denote by $m(\theta)$ the
martingale defined by the dynamics:
\begin{equation}\label{m t}\di m_t(\theta)=-\int_{\R^{d}}\sigma_t(\theta,\xi)Y^{G}(\di t,\di\xi)
-\int_{\R^{d}}\gamma_{t-}(\theta,\xi)e^{-I_{\gamma}(t-,\theta,\xi)}Y^{P}(\di t,\di\xi),\quad m_0(\theta)=0.\end{equation}
Observe that the following relation holds
\[M_t(\theta)=\int_0^{\theta}m_t(u)\di u.\]
We then consider the dynamics of the conditional density of default
given in \eqref{model:csp1}.
 \begin{prop}\label{lemma:dynamics-cd}
The conditional density
process $\alpha(\theta)$ admits the
dynamics:
\begin{eqnarray}\label{dynamics:density}
\di\alpha_t(\theta)&=&\alpha_{t-}(\theta) \di M_t(\theta)-S_{t-}(\theta)\di m_t(\theta)
\end{eqnarray}
or equivalently \[\frac{\di \alpha_t(\theta)}{\alpha_{t-}(\theta)}=\di M_t(\theta)-\frac{1}{\lambda_{t-}(\theta)}\di m_t(\theta).\]

\end{prop}

\proof
Keep the martingale condition ({\bf MC}) in mind. The dynamics (\ref{dynamics:density}) is derived by employing
It\^{o}'s formula to $\alpha(\theta)=\lambda(\theta)S(\theta)$ for each positive $\theta$ fixed.
\finproof

An important property in the credit analysis is the immersion
property, or the so called (H)-hypothesis, which means that an
$\mathbb F$-martingale remains a $\mathbb G$-martingale. The
(H)-hypothesis is satisfied if and only if
$\alpha_t(\theta)=\alpha_\theta(\theta)$ or equivalently
$\lambda_t(\theta)=\lambda_\theta(\theta)$ for any
$t\geqslant\theta$. In the random field setting, by
\eqref{model:levy-forward-intensity1}, this is equivalently to
\[\int_{\theta}^t\int_{\mathbb R^d}\sigma_s(\theta,\xi)Y^G(\di
s,\di\xi)=\int_{\theta}^t\int_{\mathbb
R^d}\gamma_{s-}(\theta,\xi)Y^P(\di s,\di \xi)=0\] for
$t\geqslant\theta$, or equivalently
\[\sigma_t(\theta,\xi)=0\quad\text{ and }\quad\gamma_t(\theta,\xi)=0\quad\nu(\di\xi)\text{-a.e.}\]
for $t>\theta$.
Note that the martingale condition ({\bf MC}) then implies that $\mu_t(\theta)=0$ for $t>\theta$.

We recall that the $\mathbb F$-intensity process $\lambda$ of the
default time $\tau$ coincides with the diagonal forward intensity,
i.e. $\lambda_t=\lambda_t(t)$. It is closely related to the Az\'ema
supermartingale:
\[S_t=S_t(t)=\mathbb P(\tau>t\,|\,\mathcal F_t),\]
which is also called the survival process of $\tau$.
\begin{prop}
Let $M$ be the $\mathbb F$-martingale having the dynamics
\[\di M_t=-\int_{\mathbb R^d}I_\sigma(t,t,\xi)Y^G(\di t,\di\xi)+\int_{\mathbb R^d}\Big(e^{-I_\gamma(t-,t,\xi)}-1\Big)Y^P(\di t,\di\xi).\]
Then
\[S_t=\exp\bigg(-\int_0^t\lambda_s\di s\bigg)\mathcal E(M)_t,\]
where $\mathcal E(M)$ is the Dol\'eans-Dade exponential of $M$.
\end{prop}
\proof
The Az\'ema supermartingale $S$ has a multiplicative decomposition of the form $S_t=L_t\exp(-\int_0^t\lambda_s\di s)$ (see \cite[Proposition 4.1]{El_Jeanblanc_Jiao_2010}), where $L$ is an $\mathbb F$-martingale
having the following dynamics
\[\di L_t=\exp\left(\int_0^t\lambda_s\di s\right)\di\widehat L_t,\]
with
\[\widehat L_t=-\int_0^t\alpha_t(u)-\alpha_u(u)\di u.\]
By Proposition \ref{lemma:dynamics-cd}, together with \eqref{M t theta} and \eqref{m t},
\[\di\widehat L_t=-\int_{\mathbb R^d}\int_0^t A(t,\theta,\xi)\di \theta Y^{G}(\di t,\di\xi)-\int_{\mathbb R^d}\int_0^t
B(t,\theta,\xi)\di \theta Y^P(\di t,\di \xi),\]
where
\[A(t,\theta,\xi)=-\alpha_{t-}(\theta)I_\sigma(t,\theta,\xi)+S_{t-}(
\theta)\sigma_t(\theta,\xi),\]
\[B(t,\theta,\xi)=\alpha_{t-}(\theta)(e^{-I_{\gamma}({t-},\theta,\xi)}-1)
+S_{t-}(\theta)\gamma_{t-}(\theta,\xi)e^{-I_{\gamma}(t-,\theta,\xi)}.\]
By integration by part, we obtain
\[-\int_0^t A(t,\theta,\xi)\di \theta=-S_{t-}(t)I_\sigma(t,t,\xi),\]
\[-\int_0^t B(t,\theta,\xi)\di\theta=S_{t-}(t)(e^{-I_{\gamma}({t-},t,\xi)}-1).\]
Moreover, the  Doob-Meyer decomposition of $S$ is
given by
\[S_t=1+\widehat{L}_t-\int_{0}^t\alpha_u(u)\di u,\]
which implies that
\[\frac{\di L_t}{L_{t-}}=\frac{\di\widehat{L}_t}{S_{t-}}=\frac{\di S_t}{S_{t-}}+\lambda_t\di t.\]
By \eqref{Equ:Sttheta}, one has  $S_{t-}=S_{t-}(t)$. Hence the martingale $L$ is the Dol\'eans-Dade exponential of $M$ and the assertion follows.
\finproof

\section{The pricing of defaultable bonds}\label{Sec: pricing}
In this section, we  focus on the pricing of credit derivatives.
In general, a credit sensitive contingent claim can be represented by a triplet $(C,G,R)$ (see {Bielecki and Rutkowski} \cite{Bielecki_Rutkowski_2002})
where the $\F_T$-measurable random variable $C_T$ represents the maturity payment if no default occurs before the
maturity $T$, and $G$ is an $\bF$-adapted continuous process of finite variation such that $G_0=0$ and represents
the coupon payment. Differently from the case where the default payment occurs at $\tau$ immediately, we assume that
in the economic default case, the default (or the recovery) payment takes place, after a period of legal proceedings, at the maturity date $T$ later
than the economic default date $\tau$ and admits the form $R_T(\tau)$ where $R_T(\cdot)$ is $\F_T\otimes\mathcal B(\R_+)$-measurable.

The global market information is described by the filtration
$\bG=(\G_t)_{t\geqslant 0}$, $\G_t=\F_t\vee\sigma(\tau\wedge t)$,
which is made to satisfy the usual conditions. The value at time
$t\leqslant T$ of the contingent claim $(C,G,R)$ is given by the following
$\G_t$-conditional expectation:
\begin{equation}\label{V G}
V_t=\esp_\Q\left[\big(C_T\indic_{\{\tau>T\}}+\int_t^T\indic_{\{\tau>u\}}e^{-\int_t^u
r_s\di s}\di G_s+\indic_{\{\tau\leqslant
T\}}R_T(\tau)\big)e^{-\int_t^T r_s\di s}\,\bigg|\,\G_t\right],
\end{equation}
where $\Q$ denotes a risk-neutral pricing probability measure and
the interest rate $r=(r_t;\ t\geqslant 0)$ is an $\bF$-adapted
process. The following result computes $V_t$ using
$\F_t$-conditional expectations. The first two terms result from
\cite{Bielecki_Rutkowski_2002} and the third one from
\cite{El_Jeanblanc_Jiao_2010}. With an abuse of notation, we denote
in the following the $\bF$-conditional density of $\tau$ under the
risk-neutral probability $\Q$ by $(\alpha_t(\theta),t\geqslant 0)$. The
general result on the density under a change of probability measure
is given in \cite[Theorem 6.1]{El_Jeanblanc_Jiao_2010}.

\begin{prop}We suppose that the economic default time $\tau$ admits a conditional density w.r.t. the filtration $\bF$,
denoted by $\alpha_t(\cdot)$ under the risk-neutral probability
measure $\Q$. Then the value of the credit sensitive contingent
claim $(C,G,R)$ is given by
\begin{equation}\label{V F}
\begin{split}
V_t&=\indic_{\{\tau>t\}}\frac{B_t}{S_t}\esp_\Q\left[\big(C_TS_T+\int_t^TR_T(u)\alpha_T(u)\di u\big)B_T^{-1}
+\int_t^TS_uB_u^{-1}\di G_u\,\bigg|\,\F_t\right]\\
&\quad+\indic_{\{\tau\leqslant
t\}}B_t\esp_\Q\left[R_T(\theta)\frac{\alpha_T(\theta)}{\alpha_t(\theta)}B_T^{-1}|\F_t\right]\Big|_{\theta=\tau}
\end{split}
\end{equation}where $S_t=\Q(\tau>t|\F_t)=\int_t^\infty\alpha_t(\theta)\di\theta$ and $B_t=\exp(\int_0^tr_s\di s)$.
\end{prop}
\proof The $\G_t$-measurable random variable $V_t$ can be decomposed in two parts
$V_t=\indic_{\{\tau>t\}}\overline V_t +\indic_{\{\tau\leq t\}}\widetilde V_t(\tau)$
where $\overline V_t$ is $\F_t$-measurable and $\widetilde V_t(\cdot)$ is $\F_t\otimes\mathcal B(\R_+)$-measurable.
On the set $\{\tau>t\}$, we use Jeulin-Yor's lemma (see \cite{Bielecki_Rutkowski_2002})
and the conditional density to obtain
\[\begin{split}\overline V_t&=\frac{1}{S_t}\esp_\Q\left[\big(C_T\indic_{\{\tau>T\}}+\indic_{\{t<\tau\leqslant T\}}R_T(\tau)\big)
e^{-\int_t^T r_s\di s}+\int_t^T\indic_{\{\tau>u\}}e^{-\int_t^u r_s\di
s}\di G_s\,\bigg|\,\F_t\right]\\
&=\frac{1}{S_t}\esp_\Q\left[\big(C_TS_T+\int_t^T
R_T(\theta)\alpha_T(\theta)\di\theta\big)e^{-\int_t^T r_s\di
s}+\int_t^TS_ue^{-\int_t^u r_s\di s}\di
G_s\,\bigg|\,\F_t\right].\end{split}\] On the set $\{\tau\leqslant
t\}$, by \cite[Thm 3.1]{El_Jeanblanc_Jiao_2010}, we have
\[\widetilde
V_t(\tau)=
\esp_\Q\left[R_T(\theta)\frac{\alpha_T(\theta)}{\alpha_t(\theta)}\exp\left(-\int_t^T
r_s\di s\right)\,\bigg|\,\F_t\right]\Big|_{\theta=\tau},
\]
which complete the proof.
\finproof

Note that for the pricing  of the two ``before default'' payment terms  $(C, G)$, the quantity
\[\frac{S_u}{S_t}=\exp\bigg(-\int_t^u\lambda_s\di s\bigg)\,\frac{\mathcal E(M)_u}{\mathcal E(M)_t}, \quad u>t \]
and hence the intensity $\lambda$ play an important role. However,
for the default recovery payment $R$ (which depends on $\tau$), the
``after-default'' density $\alpha_t(\theta)$ where
$t\geqslant\theta$ is needed. This point has been discussed in
\cite{El_Jeanblanc_Jiao_2010}. In the following of this paper, we
adopt the density approach for both the before-default and
after-default pricing.

We consider in particular a defaultable zero-coupon bond of maturity
$T$ with $C=1$ and $G=0$. Its price at $t\leqslant T$ is
given by
\begin{eqnarray}\label{eq:pricing}
P(t,T)=\E_\Q\left[\big(\indic_{\{\tau>T\}}+\indic_{\{\tau\leqslant
T\}}R_T(\tau)\big)\exp\left(-\int_t^Tr_s\di
s\right)\,\bigg|\,\G_t\right].
\end{eqnarray}
The following result is a direct consequence of the previous
proposition. We first introduce the following price kernels:
\begin{equation}\label{eq:price-kernel}
K_1(t,\theta)=\frac{1}{S_t}\E_{\Q}\left[\alpha_T(\theta)\exp\left(-\int_t^Tr_s\di
s\right)\,\Big|\,\F_t\right]\end{equation}
\begin{equation}\label{eq:price-kernebis}
K_2(t,\theta)=\frac{1}{\alpha_t(\theta)}\E_{\Q}\left[R_T(\theta)\alpha_T(\theta)\exp\left(-\int_t^Tr_s\di
s\right)\,\Big|\,\F_t\right],
\end{equation}
where $t\leqslant T$, and $\theta\geqslant 0$.

\begin{coro}\label{coro:pricing}
Using the conditional density of $\tau$ under $\Q$, the price
\eqref{eq:pricing} of the defaultable zero-coupon bond  at time
$t\leqslant T$ has the following representation:
\begin{eqnarray}\label{eq:price-formula}
P(t,T)&=&\indic_{\{\tau>t\}}\left[\int_T^{\infty}K_1(t,\theta)\di\theta+\int_t^TK_2(t,\theta)\frac{\alpha_t(\theta)}{S_t}\di\theta\right]
+\indic_{\{\tau\leqslant t\}}K_2(t,\tau).
\end{eqnarray}
\end{coro}

We will identify the above price kernels in the next two sections
with different settings.

\section{The first pricing kernel}
\label{sec:pring-kernel1}

In this section, we study in detail the pricing kernels \eqref{eq:price-kernel} and \eqref{eq:price-kernebis}
when the random interest rate is described as an extended Vasicek model. {We suppose in this section
the after-default recovery payment is deterministic.  The case where
the after-default recovery payment is random will be considered in the next section.}

Firstly we recall the forward intensity model \eqref{model:levy-forward-intensity0} and assume that the $\FF$-predictable random fields
$(\bmu,\bsigma,\bgamma)$ are deterministic in (\ref{model:levy-forward-intensity0}).
We then express the
instantaneous interest rate process $r=(r_t, t\geq 0)$ as
the following extended Vasicek model under the risk-neutral pricing measure $\Q$~:
\begin{eqnarray}\label{eq:interest-rate-field}
\di r_t=\kappa(\delta-r_t)\di t +\int_{\R^d}\rho_t(\xi)Y^{G}(\di t,
\di\xi) + \int_{\R^d}\phi_t(\xi)Y^{P}(\di t,\di\xi),
\end{eqnarray}
where $\kappa>0$, $\delta>0$, and $\rho_\cdot(\cdot)$ and
$\phi_\cdot(\cdot)$ are deterministic volatility functions, assumed
to belong to $\Phi_c$ and $\Psi_\nu$ respectively. In the particular
case where $d=0$, $\phi_t(\xi)\equiv0$ and the volatility function
$\rho_\cdot(\cdot)\equiv\rho>0$ is constant, 
the interest rate $r$ satisfies the classical Brownian-driven
Vasicek model:
\begin{eqnarray}\label{eq:interest-rate0}
\di  r_t= \kappa (\delta -r_t) \di  t+ \rho \di  W_t,
\end{eqnarray}
where $W$ is a standard Brownian motion.

Similarly to the solution form of the Ornstein-Uhlenbeck stochastic differential equation, the
extended Vasicek model (\ref{eq:interest-rate-field}) also admits an
explicit expression as follows:
\begin{equation}\label{eq:ou-solution}
\begin{split}
r_t=r_0e^{-\kappa t}+\delta&(1-e^{-\kappa
t})+\int_0^t\int_{\R^d}e^{-\kappa(t-u)}\rho_u(\xi)Y^G(\di u,\di\xi)\\
& + \int_0^t\int_{\R^d}e^{-\kappa(t-u)}\phi_u(\xi)Y^P(\di u,\di\xi),\end{split}
\end{equation}
where $r_0>0$ denotes the deterministic initial interest rate
value. 

Next we compute the first pricing kernel in (\ref{eq:price-kernel}).
For $\theta\geqslant 0$, we introduce the following
integro-differential operator $\boldsymbol{A}_\theta$ acting on
functions with three variables $t$, $x$ and $y$ which are
differential in $t$ and second-order differentiable in $(x,y)$ :
\begin{equation}\label{Equ:diff}\begin{split}\boldsymbol{A}_{\theta}K(t,x,y)&
=\kappa(\widehat{\delta}_t(\theta)-x)\frac{\partial K}{\partial x}(t,x,y)+a(t,\theta)\frac{\partial K}{\partial y}(t,x,y)+a_{11}(t)\frac{\partial^2K}{\partial x^2}(t,x,y)\\
&\quad+a_{22}(t,\theta)\frac{\partial^2K}{\partial y^2}(t,x,y)+
a_{12}(t,\theta)\frac{\partial^2K}{\partial x\partial y}(t,x,y)\\
&\quad+\int_{\mathbb R^d}\Big[K(t,x+\phi_t(\xi),y+\gamma_t(\theta,\xi))-K(t,x,y)\\
&\qquad\qquad-\phi_t(\xi)\frac{\partial K}{\partial x}(t,x,y)-
\gamma_t(\theta,\xi)\frac{\partial K}{\partial y}(t,x,y)\Big]\nu(\di\xi),
\end{split}\end{equation}
where
\begin{gather*}\label{notation A}\widehat{\delta}_t(\theta)=\delta+\kappa^{-1}\int_{\R^{2d}}\rho_t(\xi)I_\sigma(t,\theta,\xi)c(\zeta-\xi)\di\zeta
\di\xi+\kappa^{-1}\int_{\R^d}\phi_t(\xi)(e^{-I_\gamma(t,\theta,\xi)}-1)\nu(\di\xi),\\
a(t,\theta)=\mu_t(\theta)-\int_{\mathbb R^d}\sigma_t(\theta,\zeta)I_\sigma(t,\theta,\xi)c(\zeta-\xi)\di\zeta\di\xi
-\int_{\mathbb R^d}\gamma_{t^-}(\theta,\zeta)(1-e^{-I_\gamma(t,\theta,\xi)})\nu(\di\xi),\\
a_{11}(t)=\frac12\int_{\mathbb R^{2d}}\rho_t(\xi_1)\rho_t(\xi_2)c(\xi_1-\xi_2)\di\xi_1\di\xi_2,\\
a_{22}(t,\theta)=\frac 12\int_{\mathbb R^{2d}}\sigma_t(\theta,\xi_1)\sigma_t(\theta,\xi_2)c(\xi_1-\xi_2)\di\xi_1\di\xi_2,\\
a_{12}(t,\theta)=\int_{\mathbb R^{2d}}\sigma_t(\theta,\xi_1)\rho_t(\xi_2)c(\xi_1-\xi_2)\di\xi_1\di\xi_2.
\end{gather*}

\begin{rem}\label{rem:a_22=0}
Recall the martingale condition ({\bf MC}) given by \eqref{eq:S-martingale-condition} which has been assumed throughout the paper.
We have the coefficient $a(t,\theta)=0$ for the partial derivative $\frac{\partial K}{\partial y}$ under ({\bf MC}).
\end{rem}

We introduce the following assumptions where we have fixed
$\theta\geqslant 0$.
\begin{assumption}\label{Asump}
\begin{enumerate}[(1)]
\item There exists $q\in(0,1)$ such that
\begin{enumerate}[(i)]
\item the functions $a_{11}(\cdot)$, $a_{22}(\cdot,\theta)$ and $a_{12}(\cdot,\theta)$ are $\frac{q}{2}$-Lipschitz 
on $[0,T]$,
\item there exists a Borel function $J_q$ on $\mathbb R^d$ (which could depend on $\theta$) such that
\[\max\big\{|\phi_{t}(\xi)-\phi_s(\xi)|,|\gamma_t(\theta,\xi)-\gamma_s(\theta,\xi)|\big\}
\leq J_q(\xi)|t-s|^{q/2}\]
and
\[\int_{\mathbb R^d}\frac{J_q(\xi)^2}{1+J_q(\xi)}\nu(\di\xi)<\infty.\]
\end{enumerate}
\item $|\phi_t(\xi)|$ and $|\gamma_t(\theta,\xi)|$ are uniformly bounded from above by a Borel function $J_0(\xi)$ such that
    \[\int_{\mathbb R^d}\frac{J_0(\xi)^2}{1+J_0(\xi)}\nu(\di \xi)<+\infty.\]
\item There exists a constant $\beta(\theta)>0$ such that, for any $(x,y)\in\mathbb R^2$ and any $t\in[0,T]$, one has
    \[a_{11}(t)x^2+2a_{12}(t,\theta)xy+a_{22}(t,\theta)y^2\geqslant \beta(\theta)(x^2+y^2). \]
\end{enumerate}
\end{assumption}

Then we have the main result of this section.
\begin{thm}\label{thm:correlate-r}
Let $\theta\geqslant 0$ be fixed.  Under Assumption {\rm\ref{Asump}}, the
Cauchy problem
\begin{equation}\label{Equ:cauchy}\frac{\partial K}{\partial t}(t,x,y)-xK(t,x,y)+\boldsymbol{A}_\theta K(t,x,y)=0,
\quad K(T,x,y)=y\end{equation} has a unique solution $\breve{K}$,
where the integro-differential operator $\boldsymbol{A}_\theta$ is
defined in \eqref{Equ:diff}. Moreover, the following equality holds
\begin{eqnarray}\label{eq:price-kernel2}
\E_\Q\left[\alpha_T(\theta)\exp\left(-\int_t^Tr_s\di
s\right)\,\bigg|\,\F_t\right]=S_t(\theta)\breve{K}(t,r_t,\lambda_t(\theta)),
\end{eqnarray}
where $S_t(\theta)=\Q(\tau>\theta|\F_t)$ is CSP and
$\lambda_t(\theta)$ is the corresponding forward intensity under the
pricing measure $\Q$.
\end{thm}

\noindent{\it Proof.}
The first assertion comes from a general result of Garroni and Menaldi \cite[Theorem II.3.1]{Garroni_Menaldi}. Let $q$ be as in Assumption \ref{Asump}. We shall actually prove that the Cauchy problem \eqref{Equ:cauchy} with a terminal condition\footnote{The expression $C^q(\mathbb R^2)$ denotes the vector space of all bounded functions $f$ on $\mathbb R^2$ which are H\"older continuous of order $q$ (namely, such that $\|f\|_{\sup}+\|f\|_q<+\infty$), where
\[\|f\|_q:=\sup_{\begin{subarray}{c}
z,w\in\mathbb R^2\\
z\neq w
\end{subarray}}\frac{|h(z)-h(w)|}{|z-w|^q}.\]} $K(T,\cdot,\cdot)=\psi\in C^q(\mathbb R^2)$ has a unique solution in the H\"older space $C^{1+\frac{q}{2},2+q}([0,T]\times\mathbb R^2)$ by constructing a contractible operator. The case of \eqref{Equ:cauchy} with unbounded terminal function $\varphi(t,x,y)=y$ will be treated by taking limits. We recall that $C^{1+\frac{q}{2},2+q}([0,T]\times\mathbb R^2)$ denotes the vector subspace of\footnote{The expression $C^{1,2}([0,T]\times\mathbb R^2)$ denotes the vector space of all continuous functions $f$ on $[0,T]\times\mathbb R^2$ such that
\[\|f\|_{1,2}:=\sum_{a+b+2c\leq 2}\|\partial_t^c\partial_x^a\partial_y^bf\|_{\sup}<+\infty.\]
The vector space $C^{1,2}([0,T]\times\mathbb R^2)$ together with the norm $\|\cdot\|_{1,2}$ form a Banach space.
} $C^{1,2}([0,T]\times\mathbb R^2)$ of functions $f$ such that
\[\|f\|_{1+\frac{q}{2},2+q}
:=\|f\|_{1,2}+\sum_{1\leq a+b+2c\leq 2}\langle\partial_t^c\partial_x^a\partial y^b\rangle_{t,\frac 12(q+a+b+2c-1)}+\sum_{a+b+2c=2}\langle\partial_t^c\partial_x^a\partial y^b\rangle_{(x,y),q}<+\infty,\]
where for any function $g:[0,T]\times\mathbb R^2\rightarrow\mathbb R$ and any $\beta\in(0,1)$,
\[
\langle g\rangle_{t,\beta}:=\sup_{z\in\mathbb R^2}\sup_{\begin{subarray}{c}
s,t\in[0,T]\\
s\neq t
\end{subarray}}\frac{|g(s,z)-g(t,z)|}{|s-t|^\beta},\;
\langle g\rangle_{(x,y),\beta}:=\sum_{t\in[0,T]}\sup_{\begin{subarray}{c}z,w\in\mathbb R^2\\
z\neq w
\end{subarray}}\frac{|g(t,z)-g(t,w)|}{|z-w|^\beta}.\]
The vector space $C^{1+\frac{q}{2},2+q}([0,T]\times\mathbb R^2)$ together with the norm $\|\cdot\|_{1+\frac{q}{2},2+q}$ form a Banach space.

Let $\boldsymbol{I}_\theta$ be the integro-differential operator defined as
\[(\boldsymbol{I}_{\theta}K)(t,x,y)=\int_{\mathbb R^d}\bigg[K(t,x+\phi_t(\xi),y+\gamma_t(\theta,\xi))-K(t,x,y)-\phi_t(\xi)\frac{\partial K}{\partial x}-
\gamma_t(\theta,\xi)\frac{\partial K}{\partial y}\bigg]\nu(\di\xi).\]
For $K\in C^{1+\frac{q}{2},2+q}([0,T]\times\mathbb R^2)$ and $\psi\in C^q(\mathbb R^2)$, let $\Theta_\psi(K)$ be the unique solution of the Cauchy problem
\begin{equation}
\label{Equ:diff_auxi}
\frac{\partial F}{\partial t}-xF +\widehat{\boldsymbol{A}}_\theta(F)=\boldsymbol{I}_\theta(K),\quad
F(T,x,y)=\psi(x,y),
\end{equation}
where $\widehat{\boldsymbol{A}}_{\theta}$ denotes the differential operator
\[a_{11}(t)\frac{\partial^2 }{\partial x^2}+a_{12}(t,\theta)\frac{\partial^2}{\partial x\partial y}+a_{22}(t,\theta)\frac{\partial^2}{\partial y^2}+\kappa(\widehat{\delta}_t(\theta)-x)\frac{\partial }{\partial x}+a(t,\theta)\frac{\partial}{\partial y}.\]

Denote by $C^{\frac{q}{2},q}([0,T]\times\R^2)$ the vector space of functions $f$ on $[0,T]\times\mathbb R^2$ such that
\[\|f\|_{\frac{q}{2},q}:=\|f\|_{\sup}+\sup_{(x,y)\in\mathbb R^2}\|f(\cdot,x,y)\|_{\frac{q}{2}}+\sup_{t\in[0,T]}\|f(t,\cdot,\cdot)\|_{q}<+\infty\]
which is a Banach space with respect to the norm
$\|\cdot\|_{\frac{q}{2},q}$. Since $K\in
C^{1+\frac{q}{2},2+q}([0,T]\times\mathbb R^2)$, by the
Assumption \ref{Asump} (1.ii) and (2), we obtain that
$\boldsymbol{I}_\theta(K)\in
C^{\frac{q}{2},q}([0,T]\times\R^2)$ (see \cite[Lemma
II.1.5]{Garroni_Menaldi}). Therefore the existence and uniqueness of
the solution $\Theta_\psi(K)\in
C^{1+\frac{q}{2},2+q}([0,T]\times\mathbb R^2)$ to
\eqref{Equ:diff} comes from the classical theory of parabolic
partial differential equations (e.g. \cite{Friedman69}). Moreover,
the solution verifies the following H\"older estimate (\cite[Theorem
I.2.1]{Garroni_Menaldi})
\begin{equation}\label{Equ:Holder}\|\Theta_{\psi}(K_1)-\Theta_{\psi}(K_2)\|_{{1+\frac{q}{2},2+q}}\leqslant C_1\|\boldsymbol{I}_\theta(K_1-K_2)\|_{{\frac{q}{2},q}}\end{equation}
which holds for all $K_1,K_2\in C^{1+\frac{q}{2},2+q}([0,T]\times\mathbb R^2)$ such that $K_1(T,\cdot,\cdot)=K_2(T,\cdot,\cdot)=\psi$, where $C_1$ is a constant independent of $\psi$.

For arbitrary $\varepsilon>0$, the following estimate holds for any $K\in C^{1+\frac{q}{2},2+q}([0,T]\times\mathbb R^2)$ (see \cite[Lemma II.1.5]{Garroni_Menaldi})
\begin{equation}\label{Equ:IthetaK}
\|\boldsymbol{I}_\theta(K)\|_{_{\frac{q}{2},q}}
\leqslant \varepsilon\|\nabla_{(x,y)}^2K\|_{_{\frac{q}{2},q}}
+ C(\varepsilon)\Big(\|K\|_{\frac{q}{2},q}+\|\nabla_{(x,y)}(K)\|_{\frac{q}{2},q}\Big),\end{equation}
where the constant $C(\varepsilon)$ only depends on $\varepsilon$.
Denote by $C^q_\psi$ the subset of functions in $C^{1+\frac{q}{2},2+q}([0,T]\times\mathbb R^2)$ whose restriction on $\{T\}\times\mathbb R^2$ coincides with $\psi$.
By choosing
$\varepsilon>0$ small enough, we obtain from \eqref{Equ:Holder} and \eqref{Equ:IthetaK} that $\Theta_{\psi}$ is a contracting operator on the complete metric space $C^q_\psi$, provided that $T$ is sufficiently small. Hence for sufficiently small $T$, the operator $\Theta_{\psi}$ has a unique fixed point and therefore the Cauchy problem
\[\frac{\partial K}{\partial t}(t,x,y)-xK(t,x,y)+\boldsymbol{A}_\theta K(t,x,y)=0,
\quad K(T,x,y)=\psi(x,y)\]has a unique solution.
For general $T$, it suffices to divide $[0,T]$ into a finite union of small intervals and resolve the Cauchy problem progressively.

For the terminal function $\varphi(x,y)=y$, we can take, for each integer $n\geq 1$, a function $\psi_{n}\in C_0^\infty(\R^2)$ which coincides with $\varphi$ on the ball $B_n$ of radius $n$ centered at $(0,0)$. For any $n\geqslant 1$, let $K_n$ be the unique solution of the equation \eqref{Equ:cauchy} with terminal condition $K_n(T,x,y)=\psi_n(x,y)$. By a maximum principle for the equation \eqref{Equ:cauchy} (see  \cite[Theorem II.2.15]{Garroni_Menaldi}), for $n\geqslant m$, $K_n$ coincides with $K_m$ on the ball $B_m$. By taking $\breve{K}=K_n$ on $[0,T]\times B_n$ we obtain a global solution to the Cauchy problem \eqref{Equ:cauchy}. The uniqueness of $\breve{K}$ also results from the maximum principle.

We now prove the second assertion that $\breve{K}$ satisfies the equality \eqref{eq:price-kernel2}. To this end, we compute the denominator of the pricing kernel \eqref{eq:price-kernel}  by
introducing a change of probability measure:
\begin{eqnarray}\label{def:radon_Nik}
\frac{\di\Q^\theta}{\di\Q}\Big|_{\F_t}=\frac{S_t(\theta)}{S_0(\theta)}.
\end{eqnarray}
By Bayes' formula and \eqref{eq:relation}, we have
\[\E_\Q\left[\alpha_T(\theta)\exp\left(-\int_t^Tr_s\di
s\right)\,\bigg|\,\F_t\right]=S_t(\theta)\mathbb
E_{\Q^\theta}\left[\lambda_T(\theta) \exp\left(-\int_t^Tr_s\di
s\right)\,\bigg|\,\mathcal F_t\right].\] Note that, by Girsanov's
theorem (see
\cite[Theorem 3.3]{Bjork_Kabanov_Runggaldier}), 
under the probability measure $\Q^\theta$,
$$
\widehat{Y}^G(\di t,\di\xi):={Y}^G(\di
t,\di\xi)+\left(\int_{\R^d}I_\sigma(t,\theta,\xi)c(\zeta-\xi)\di\zeta\right)
\di\xi\di t
$$
defines a Gaussian field with correlated kernel $c$ on $\R^d$, and
$$
\widehat{Y}^P(\di t,\di\xi):={Y}^P(\di
t,\di\xi)+(1-e^{-I_\gamma(t,\theta,\xi)})\nu(\di\xi)\di t
$$
defines a compensated Poisson random measure with predictable compensator $e^{-I_\gamma(t,\theta,\xi)}\nu(\di\xi)\di t$.
Then the dynamics \eqref{eq:interest-rate-field} of the interest rate $r$ can be rewritten as
\begin{eqnarray*}
\di r_t=\kappa(\widehat{\delta}_t(\theta)-r_t)\di
t+\int_{\R^d}\rho_t(\xi)\widehat{Y}^{G}(\di t,\di\xi)
+\int_{\R^d}\phi_t(\xi)\widehat{Y}^{P}(\di t,\di\xi),
\end{eqnarray*}
where
\[\widehat{\delta}_t(\theta)=\delta+\kappa^{-1}\int_{\R^{2d}}\rho_t(\xi)I_\sigma(t,\theta,\xi)c(\zeta-\xi)\di\zeta
\di\xi+\kappa^{-1}\int_{\R^d}\phi_t(\xi)(e^{-I_\gamma(t,\theta,\xi)}-1)\nu(\di\xi),\]
and the dynamics \eqref{model:levy-forward-intensity0}
 of the forward intensity rate can be rewritten as
\[\di\lambda_t(\theta)=\widehat{\mu}_t(\theta)\di t+\int_{\mathbb R^d}\sigma_t(\theta,\xi)\widehat Y^G(\di t,\di\xi)+\int_{\mathbb R^d}\gamma_{t-}(\theta,\xi)\widehat Y^P(\di t,\di\xi)\]
where
\[\widehat{\mu}_t(\theta)=\mu_t(\theta)-\int_{\mathbb R^d}\sigma_t(\theta,\zeta)I_\sigma(t,\theta,\xi)c(\zeta-\xi)\di\zeta\di\xi
-\int_{\mathbb R^d}\gamma_{t-}(\theta,\xi)(1-e^{-I_\gamma(t,\theta,\xi)})\nu(\di\xi).\]
Note that the forward intensity process $\lambda(\theta)$ is a $(\Q^\theta,\FF)$-martingale for each
$\theta$ fixed. Assume that
$$
\E_{\Q^\theta}\left[\lambda_T(\theta)\exp\left(-\int_t^Tr_s\di s\right)\,\bigg|\,\F_t\right]=K(t,r_t,\lambda_t(\theta)),
$$
where the function $K(t,x,y)$ is sufficiently regular. Then It\^o's formula applied to the $(\mathbb Q^\theta,\mathbb F)$-martingale
\[\exp\left(-\int_0^tr_s\di s\right)K(t,r_t,\lambda_t(\theta))\]
yields
\[\begin{split}&-r_tK(t,r_t,\lambda_t(\theta))+\frac{\partial K}{\partial t}(t,r_t,\lambda_t(\theta))+\kappa(\widehat{\delta}_t(\theta)-r_t)\frac{\partial K}{\partial x}(t,r_t,\lambda_t(\theta))+\widehat{\mu}_t(\theta)\frac{\partial K}{\partial y}(t,r_t,\lambda_t(\theta))\\
&+a_{11}(t)
\frac{\partial^2 K}{\partial x^2}(t,r_t,\lambda_t(\theta))+
a_{22}(t,\theta)\frac{\partial ^2K}{\partial y^2}(t,r_t,\lambda_t(\theta))+a_{12}(t,\theta)\frac{\partial^2K}{\partial x\partial y}(t,r_t,\lambda_t(\theta))\\
&+\int_{\mathbb R^d}\Big[K(t,r_t+\phi_t(\xi),\lambda_t(\theta)+\gamma_t(\theta,\xi))-K(t,r_t,\lambda_t(\theta))\\
&\qquad\quad-\phi_t(\xi)\frac{\partial K}{\partial x}(t,r_t,\lambda_t(\theta))-
\gamma_t(\theta,\xi)\frac{\partial K}{\partial y}(t,r_t,\lambda_t(\theta))\Big]\nu(\di\xi)=0.\\
\end{split}\]
Conversely, if $\breve{K}$ is the solution to
\[\frac{\partial K}{\partial t}-xK+\boldsymbol{A}_\theta K=0,\quad K(T,x,y)=y,\]
then one has
\[\mathbb E_{\mathbb Q}\left[\alpha_T(\theta)\exp\left(-\int_t^Tr_s\di s\right)\,\bigg|\,\mathcal F_t\right]=S_t(\theta)\breve{K}(t,r_t,\lambda_t(\theta)).\]
Thus we complete the proof of the theorem.
\finproof

Accordingly the pricing kernels \eqref{eq:price-kernel} and \eqref{eq:price-kernebis} are given by
\begin{eqnarray}K_1(t,\theta)&=&\frac{S_t(\theta)}{S_t}\breve{K}(t,r_t,\lambda_t(\theta))\\
K_2(t,\theta)&=&\frac{S_t(\theta)}{\alpha_t(\theta)}R_T(\theta)\breve{K}(t,r_t,\lambda_t(\theta))\end{eqnarray}
where $t\leqslant T$ and $\theta\geqslant 0$. By Corollary
\ref{coro:pricing}, we obtain immediately the pricing formula for
the defaultable zero-coupon bond.

\begin{rem}
Concerning the pricing kernel at the left side of the equality \eqref{eq:price-kernel2}, one possible alternative way is to solve it directly by using the dynamics of the density $\alpha_t(\theta)$. However, in view of \eqref{dynamics:density}, the corresponding solution $K(t,r_t,S_t(\theta),\lambda_t(\theta))$ will include three variables apart from time variable. The main advantage of the change of probability method \eqref{def:radon_Nik} is that we obtain the solution function in the form $K(t,r_t,S_t(\theta),\lambda_t(\theta))=S_t(\theta)\check{K}(t,r_t,\lambda_t(\theta))$. This indeed decreases the dimension of variables for our pricing kernel function and is important in the numerical computation.

\end{rem}

\begin{rem}\label{rem:5.5}
If the interest rate $r$ is independent of the forward intensity,
hence independent of the density, then the computation of the
pricing kernels is easier. Denote by $B(t,T)$ the price of the
standard zero-coupon bond, i.e. $B(t,T)=\esp_{\Q}[\exp(-\int_t^T
r_s\di s)|\F_t]$. Recall that we have assumed the recovery rate
deterministic in this section.  Then
\[K_1(t,\theta)=\frac{\alpha_t(\theta)B(t,T)}{S_t},\quad K_2(t,\theta)=R_T(\theta)B(t,T)\]which implies that
the time-$t$ value
\eqref{eq:price-formula} of defaultable zero-coupon bond has the following representation:
\begin{gather}\label{eqcoro:price-formula}
\frac{P(t,T)}{B(t,T)}=\indic_{\{\tau>t\}}\left(
1-\frac{\int_t^T(1-R_T(\theta))\alpha_t(\theta)\di\theta}{S_t}\right)
+\indic_{\{\tau\leqslant t\}}R_T(\tau).
\end{gather}
This quantity serves to measure the default risk including both the default probability and the loss given default. We also notice in \eqref{eqcoro:price-formula} that the recovery corresponds to a ``recovery of face value''
since it can be written as the quotient between the defaultable bond and an equivalent default-free bond.

\end{rem}

\begin{rem}
The zero coupon price $B(t,T):=\mathbb E_{\mathbb
Q}\Big[\exp\Big(-\int_t^Tr_s\di s\Big)\,\Big|\,\mathcal F_t\Big]$
can be given in the form $K(t,r_t)$ where $K(\cdot,\cdot)$ is the
unique solution to the following integro-differential equation:
\[-xK+\frac{\partial K}{\partial t}+\kappa(\delta-x)\frac{\partial K}{\partial x}
+a_{11}(t)\frac{\partial^2K}{\partial x^2}+\int_{\mathbb R^d}\Big[K(t,x+\phi_t(\xi))-K(t,x)-\phi_t(\xi)\frac{\partial K}{\partial x}(t,x)\Big]\nu(\di \xi)=0\]
with the terminal condition $K(T,x)=1$. If there is no jumps in $r_t$, i.e., $\phi_t(\xi)\equiv 0$, then the above equation becomes
\[-xK+\frac{\partial K}{\partial t}+\kappa(\delta-x)\frac{\partial K}{\partial x}+a_{11}(t)\frac{\partial^2K}{\partial x^2}=0.\]
Its unique solution is
\[\widehat K(t,x)=\exp\bigg(\frac{1-e^{-\kappa(T-t)}}{\kappa}(\delta-x)-\delta(T-t)+\int_t^Ta_{11}(u)\Big(\frac{1-e^{-\kappa(T-u)}}{\kappa^2}\Big)^2\di u\bigg),\]
where $a_{11}(t)$ is given in \eqref{Equ:diff}. Thus we obtain the
following equality $B(t,T)=\widehat K(t,r_t)$,
which is similar to the classical case.
\end{rem}

\section{Random recovery rate and the second pricing kernel}
\label{sec:pring-kernel2}
In this section, we consider the general case for the pricing kernel \eqref{eq:price-kernebis}, where the after-default recovery payment is random as an extension to the previous section.

Bakshi et al. \cite{Bakshi_Madan_Zhang} assumed that the recovery
rate is related to the underlying intensity as the following form:
$R_t=w_0+w_1e^{-\lambda_t}$,
$w_0,w_1\geqslant 0$,
$w_0+w_1\leqslant 1$ and $\lambda$ is the intensity process of
default. In a similar manner, we assume that $R_T(\theta)$ is of the
form
\begin{equation}\label{Equ:modelRT}R_T(\theta)=w_0+w_1e^{-f(\lambda_T(\theta))},\ \ \ \theta\geqslant0\end{equation}
where $\lambda_T(\theta)$ is the forward intensity implied by
\eqref{def:forward-intensity} under the pricing measure $\Q$, $w_0$,
$w_1$ satisfy the same condition as above and $f$ is a non-negative
function which is locally H\"older continuous of positive order.

\begin{prop}\label{prop:sec6}
Let $\theta\geqslant 0$ be fixed. Under the Assumption
{\rm\ref{Asump}}, the pricing kernel
\eqref{eq:price-kernebis}  is given by
\begin{equation}\label{Equ:K2}K_2(t,\theta)=\frac{w_0}{\lambda_t(\theta)}\breve{K}(t,r_t,\lambda_t(\theta))+\frac{w_1}{\lambda_t(\theta)}\widetilde{K}(t,r_t,\lambda_t(\theta)),\end{equation}
where $\breve{K}$ and $\widetilde{K}$ are  respectively solutions to
the partial integro-differential equation:
\begin{equation}\label{Equ:PIDE}\frac{\partial K}{\partial t}(t,x,y)-xK(t,x,y)+\boldsymbol{A}_{\theta}K(t,x,y)=0\end{equation}
under the terminal conditions $\breve{K}(T,x,y)=y$ and $\widetilde K(T,x,y)=ye^{-f(y)}$.
\end{prop}
\proof
Similarly to Theorem \ref{Equ:cauchy}, the equation \eqref{Equ:PIDE} with the terminal condition $K(T,x,y)=ye^{-f(y)}$ admits a unique solution $\widetilde K$. Moreover, by a change of probability measure we obtain
\[\mathbb E_{\mathbb Q}\bigg[
\alpha_T(\theta)e^{-f(\lambda_T(\theta))}\exp\bigg(-\int_t^Tr_s\di s\bigg)\,\bigg|\,\mathcal F_t\bigg]=S_t(\theta)\widetilde K(t,r_t,\lambda_t(\theta)).\]
Hence the formula \eqref{Equ:K2} follows from the following relation (see \eqref{eq:price-kernel}, \eqref{eq:price-kernebis} and \eqref{Equ:modelRT}) :
\[K_2(t,\theta)=\frac{w_0S_t}{\alpha_t(\theta)}K_1(t,\theta)+
\frac{w_1}{\alpha_t(\theta)}\mathbb
E\bigg[\alpha_T(\theta)e^{-f(\lambda_T(\theta))}\exp\bigg(-\int_t^Tr_s
\di s\bigg)\,\bigg|\,\mathcal F_t\bigg],\] where $K_1(t,\theta)$ is
the first price kernel \eqref{eq:price-kernel}. \finproof

\begin{coro}
Under the Assumption {\rm\ref{Asump}}, the price of the defaultable
zero-coupon bond is given by
\[\begin{split}P(t,T)=&\indic_{\{\tau>t\}}\bigg[\int_T^\infty\frac{S_t(\theta)}{S_t}\breve{K}(t,r_t,\lambda_t(\theta))\di\theta
+\int_t^T \frac{S_t(\theta)}{S_t}\Big(w_0\breve{K}(t,r_t,\lambda_t(\theta))+w_1\widetilde{K}(t,r_t,\lambda_t(\theta))\Big)\di\theta\bigg]\\
&+\indic_{\{\tau\leqslant
t\}}\frac{1}{\lambda_t(\tau)}\left[{w_0}\breve{K}(t,r_t,\lambda_t(\tau))+{w_1}\widetilde
K(t,r_t,\lambda_t(\tau))\right],
\end{split}\]
where $\breve{K}$ and $\widetilde{K}$ are given in Proposition
{\rm\ref{prop:sec6}} respectively.
\end{coro}

\section{Numerical illustrations}\label{sec:numerical}

In this section, we illustrate our previous results by numerical examples. We are particularly interested in
the contagion phenomenon. More precisely, we shall analyze in detail the the jump part in the default density dynamics  and its impact on the defaultable bond pricing.

In the numerical example, we consider the dynamics of the default density described by \eqref{dynamics:density} and  we let the martingale $m(\theta)$ be given by
\begin{equation}\label{eq:num_m}
\di m_t(\theta)=- \sigma_t(\theta)\di W_t+\int_{\R_+}\gamma_{t-}(\theta) \xi{e^{-\xi\int_0^{\theta}\gamma_{t-}(v)\mathrm{d}v}} Y^P(\di t,\di\xi),\ \ m_0(\theta)=0,\end{equation} with $W=(W_t;\ t\geqslant 0)$ being a standard Brownian motion independent of the Poisson measure $Y^P$.
Compared  with \eqref{m t}, the corrected kernel $c$ of the Gaussian field $Y^G$ is the Dirac measure and $d=1$, the volatility coefficient
$\sigma_t(\theta,\xi)=\sigma_t(\theta)$ does not depend on $\xi$ and the jump amplitude coefficient is given by $\gamma_t(\theta,\xi)=\gamma_t(\theta)\xi\1_{\{\xi>0\}}$ where $\gamma_t(\theta)>0$. Recall in addition that  $M_t(\theta)=\int_0^\theta m_t(u)\di u$ and
\[
\di\alpha_t(\theta)=\alpha_{t-}(\theta)\di M_t(\theta)-S_{t-}(\theta)\di m_t(\theta).
\]

To illustrate the impact of the jump part on the defaultable bond price $P(t,T)$ given by \eqref{eq:price-formula},
we first consider the case when the martingale $m(\theta)$ has no jumps, i.e.,
$\gamma=0$. We then include the jump part in the density dynamics. 
We use the initial default density given by $\alpha_0(\theta)=\lambda  e^{-\lambda \theta}$ with $\lambda$ being a positive constant.

In the coming tests,  we suppose that $\sigma_t(\theta)$ and $\gamma_t(\theta)$ are deterministic and we use the following forms of the coefficients and the characteristic measure in \eqref{eq:num_m},
\[
\begin{cases}
\sigma_t(\theta)=\sigma(\theta-t)^+,\ \ \sigma>0,\\
\gamma_t(\theta)=b(\theta-t)^{+},\ \ b>0,\\
\nu(\di\xi)=\frac{\zeta}{\varpi}  e^{-\xi/\varpi}\1_{\{\xi>0\}}\di\xi,\ \ \ \zeta> 0,\varpi> 0.
\end{cases}
\]
We assume that both the recovery rate $R\in[0,1]$ and
the interest rate $r$ are constants and define $B(t,T)=e^{-r(T-t)}$ for $0\leqslant t\leqslant T$. By Remark \ref{rem:5.5}, the defaultable bond price $P(t,T)$ given by \eqref{eq:price-formula} admits an explicit form. Since the quotient $P(t,T)/B(t,T)$ equals the constant $R$ on the set $\{\tau\leqslant t\}$ in this case,  we only study the pre-default part on $\{\tau>t\}$ in \eqref{eqcoro:price-formula}, which is denoted by $P(t,T)$ henceforth and is given by
\begin{eqnarray}\label{eq:indep-price-2}
P(t,T)&=&B(t,T)\left(1-(1-R)\frac{\int_t^T\alpha_t(\theta)\di\theta}
{\int_t^\infty\alpha_t(\theta)\di\theta} \right).
\end{eqnarray}
The main task is then to approximate the integral $\int_t^\infty\alpha_t(\theta)\di\theta$ by a finite sum
$\sum_{i=t/\Delta+1}^{N/\Delta} \Delta*\alpha_t(i*\Delta)$.
Here we choose $\Delta=1/100$ and $N=10/\lambda$.
We perform $10^4$ experiments to compute the $\F_t$-measurable random variable $P(t,T)$. In each
experiment, we first generate the underlying Brownian motion and the central compound Poisson process.
Then for each $\theta\in \{i\Delta; i=1,2,\dots, N/\Delta\}$, we compute $\alpha_{t_i}(\theta)$
on $\{t_i=i\Delta_t; i=1,2,\dots,t/\Delta_t\}$ with $\Delta_t=1/100$.

The preferred parameter values are as follows:
\[
t=0.5,\ T=1,\ r=0.05,\ R=0.4,\ b=1,\ \zeta=10,\ \lambda=0.1.
\]

Figure \ref{fig:price_density} plots the kernel estimations of the densities of $P(t,T)$ given by
\begin{eqnarray}\label{eq:ker-den-price}
f_P(x):= \frac{1}{k}\sum_{i=1}^k f_h(x-P_i(t,T)),
\end{eqnarray}
where $P_i(t,T)$ is the price obtained in the $i$-th experiment, $f_h(x)=\frac{1}{\sqrt{2\pi}h}\exp\left(-\frac{x^2}{2h^2}\right)$, and $h=1.06 s_k k^{-1/5}$ is the bandwidth, with $s_k$ being the sample standard deviation. From Figure \ref{fig:price_density}, we find that the existence of the  jump risk will increases the decentrality of the price. The right tail of the price distribution becomes fatter and fatter as the mean jump size $\varpi$ increasing.

\begin{center}
\makeatletter
\def\@captype{figure}
\caption{\small  The (normal) kernel estimations of the price densities for $\varpi=0, 0.0002, 0.0006, 0.001, 0.002$ and $\sigma=0.001$.\label{fig:price_density}}

\makeatother
\includegraphics[width=3.6in]{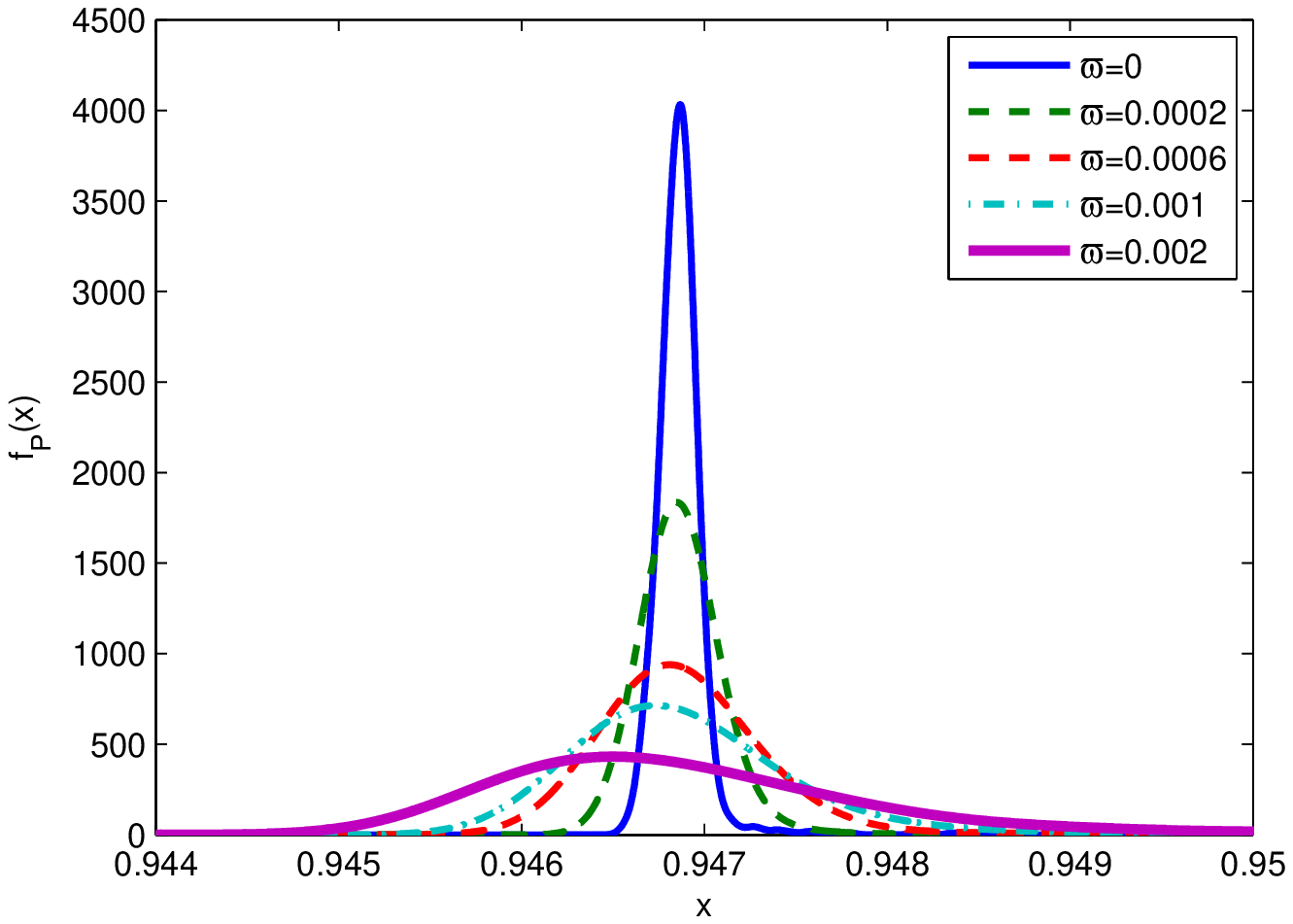}
\end{center}

\begin{center}
\makeatletter
\def\@captype{figure}
\caption{\small  $P_e(0.5,1):=\E[P(0.5,1)]$ as a function of $\varpi$ with $\lambda=0.01, 0.03, 0.1, 0.3$ and $\sigma=0.001$. \label{fig:pricevsmu}}

\makeatother
\includegraphics[width=3in]{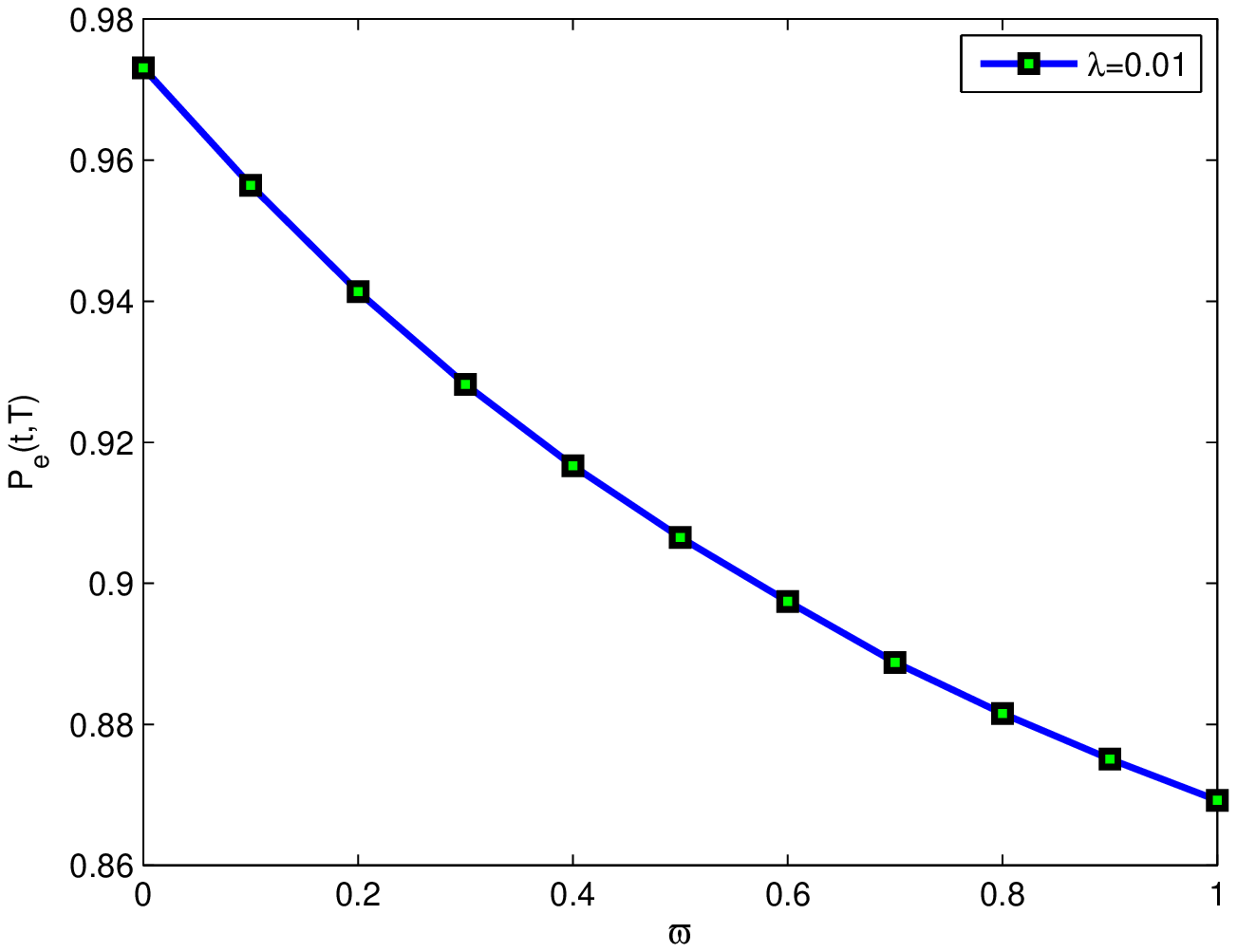}
\includegraphics[width=3in,height=2.26in]{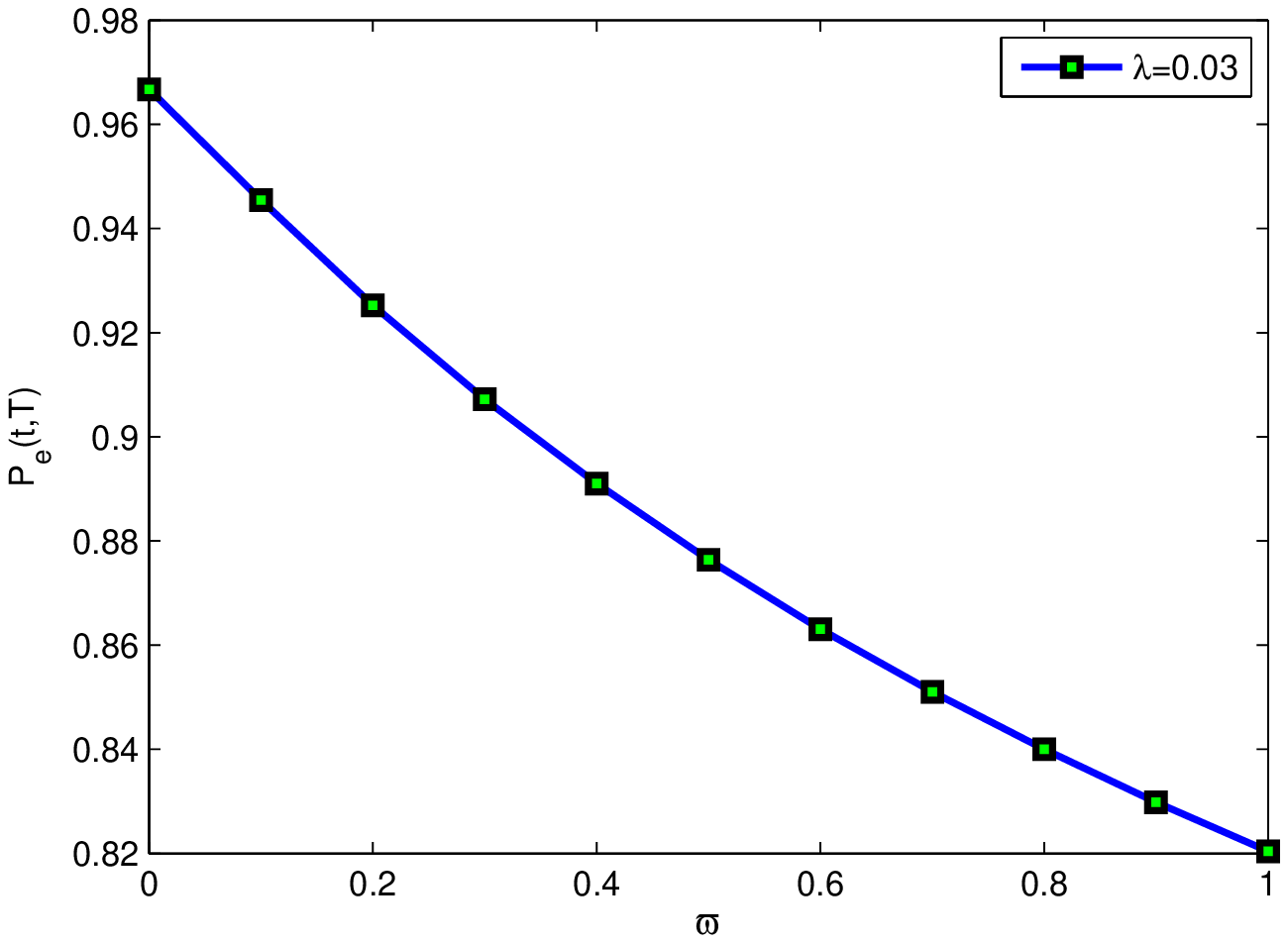}
\includegraphics[width=3in]{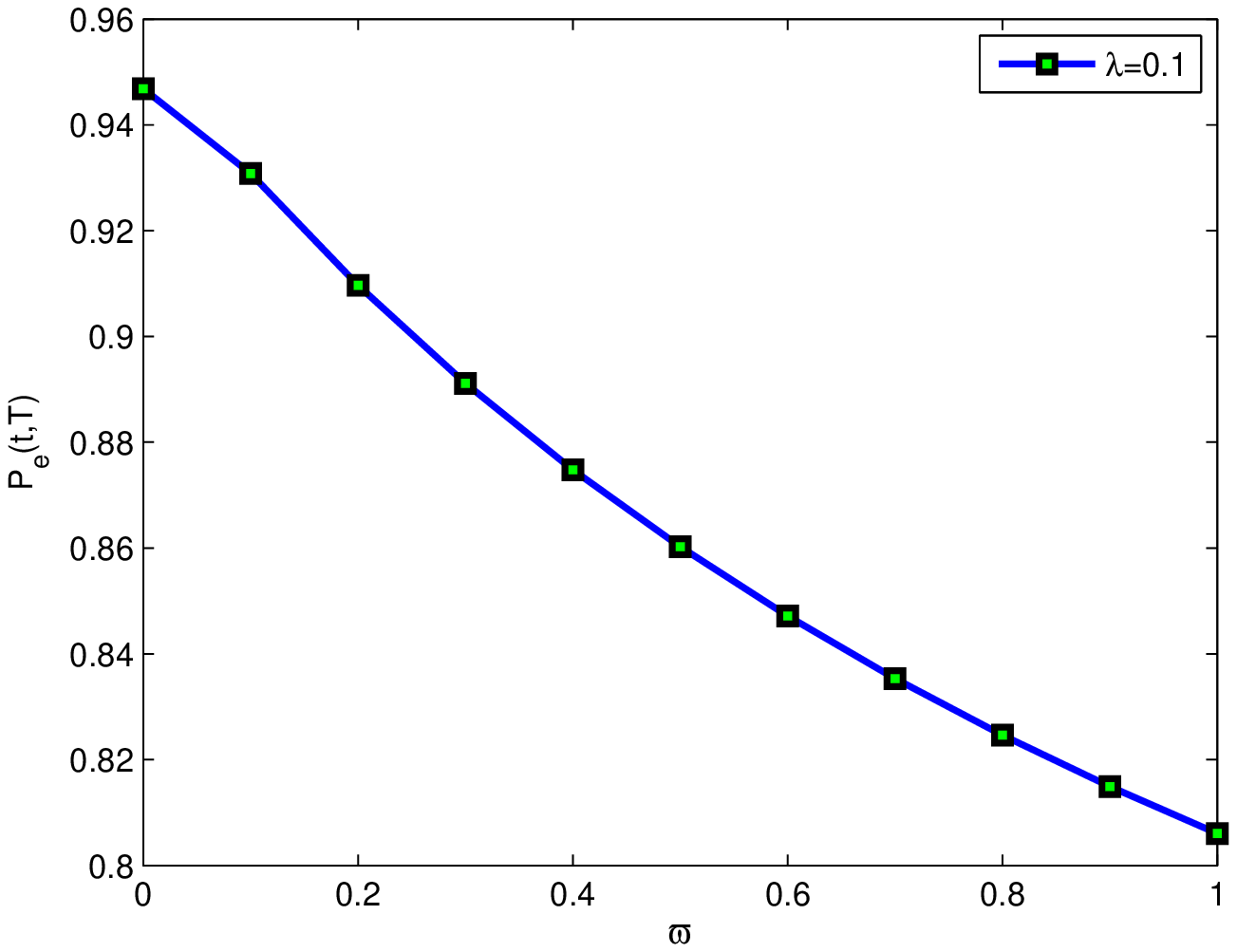}
\includegraphics[width=3in]{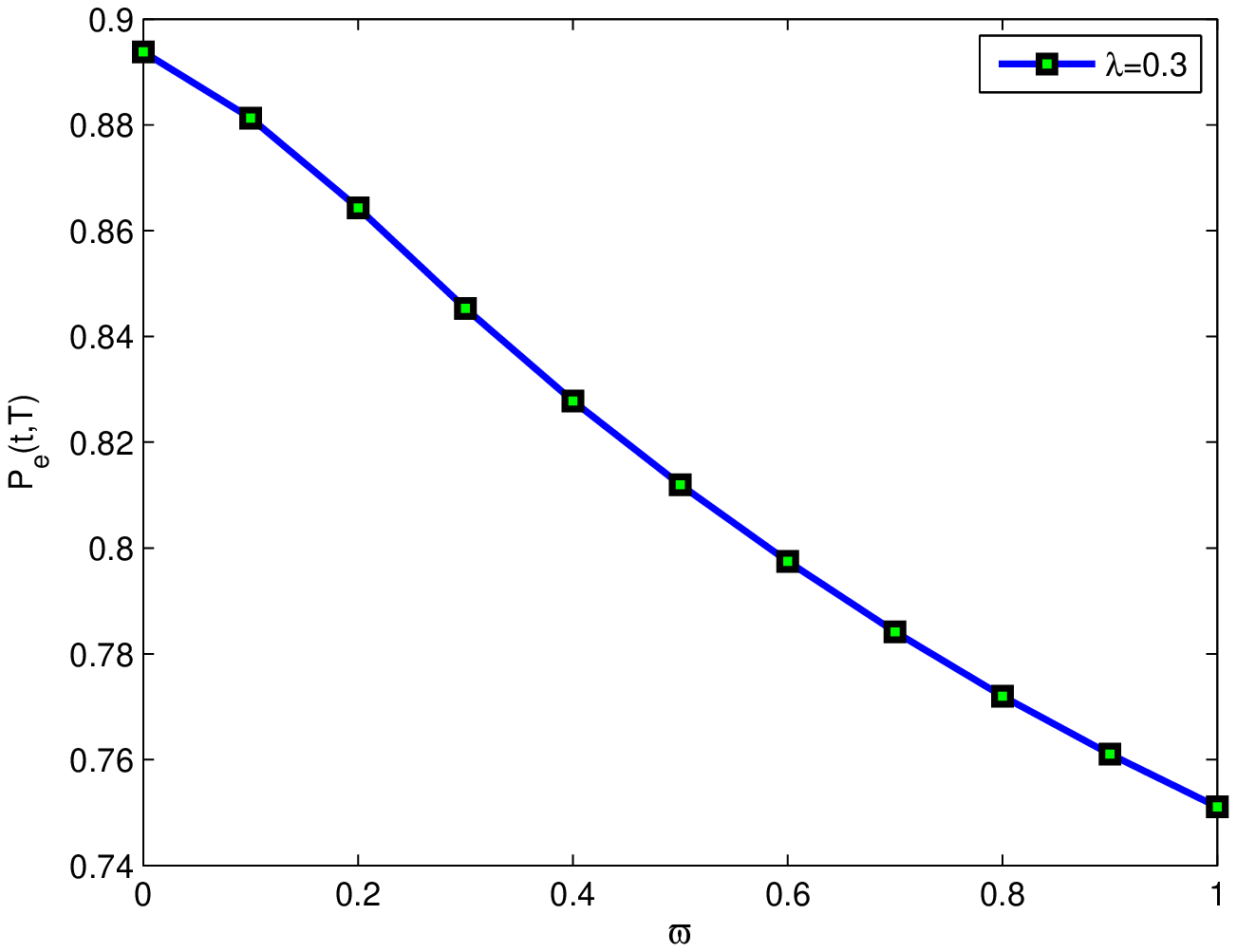}
\end{center}

Figure \ref{fig:pricevsmu} shows the mean of the price $P_e(0.5,1):=\E[P(0.5,1)]$ as a function of $\varpi$ for different values of $\lambda$. We observe that the defaultable bond price is a decreasing function of the intensity $\lambda$, and also of the mean jump size $\varpi$. Hence, when there is larger default risk of the underlying asset itself (with larger $\lambda$), the corresponding bond price is smaller. Furthermore, when there is more significant counterparty risks, that is, when there is a larger contagious jump in the density (larger $\varpi$), then the bond price will also decrease. Both observations correspond to the reality on the market.

Figure \ref{fig:pricevst} shows the mean of the price $P_e(t,1):=\E[P(t,1)]$ as a function of $t$. It is noted that the numerical illustration of the quantity $P(t,T)/B(t,T)$ discussed in Remark \ref{rem:5.5} is very similar to that of $P(t,T)$, since $B(t,T)$ here is a deterministic function $B(t,T)=e^{-r(T-t)}$ which is close to $1$. We observe similar results as in the previous test: the counterparty jump risks in the density will decrease the bond prices.



\begin{center}
\makeatletter
\def\@captype{figure}
\caption{\small $P_e(t,1):=\E[P(t,1)]$ as a function of $t$. Right hand side is the relative price $P_e(t,1):=\E[P(t,1)]/B(t,1)$.\label{fig:pricevst}}
\makeatother
\includegraphics[width=3in]{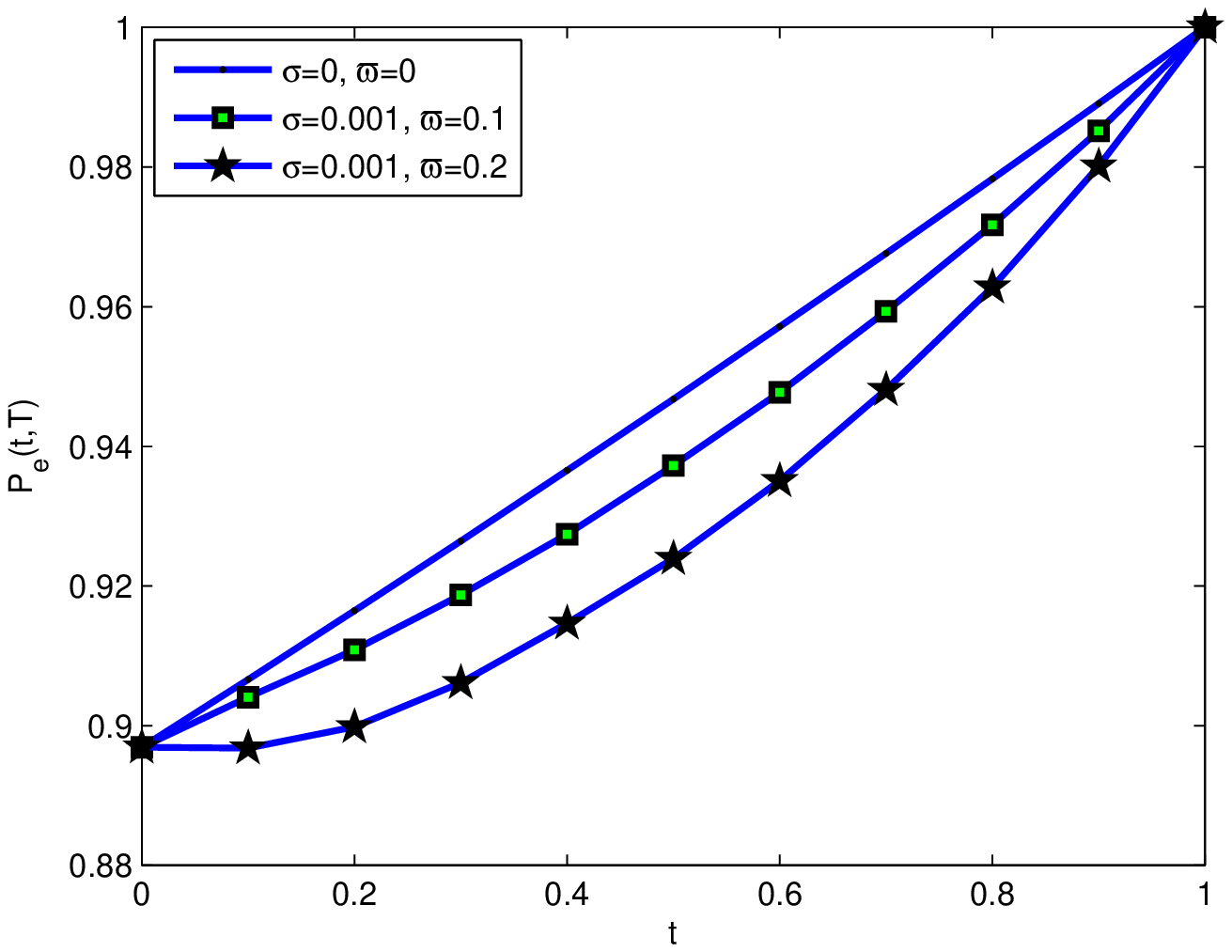}
\includegraphics[width=3in]{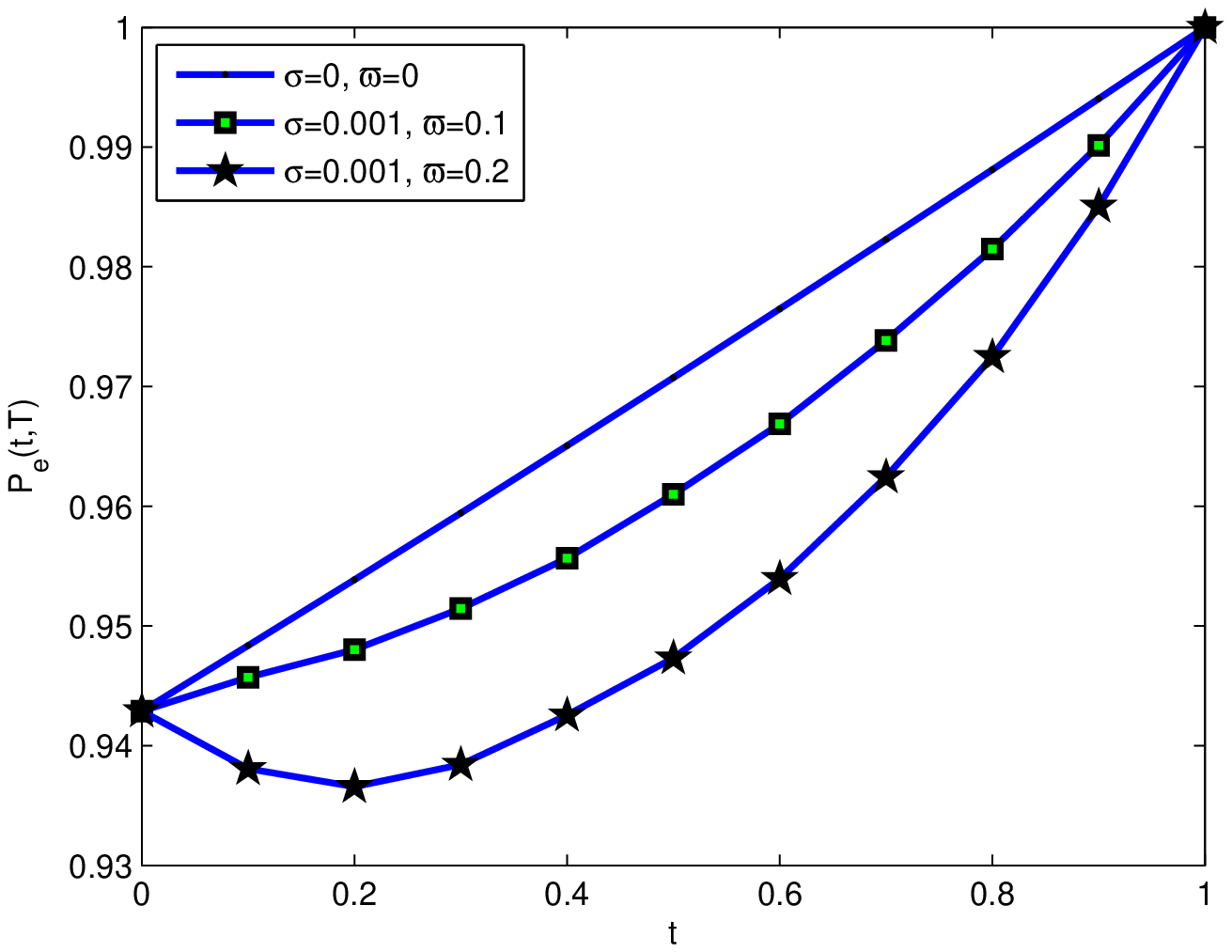}

\end{center}

In the last graph, we show the quoted bond price at the initial time $t=0$ as a function of the maturity time $T$ for different values of  intensities. Again we observe that the bond price is decreasing when there is larger default risks and for long term bonds.
\begin{center}
\makeatletter
\def\@captype{figure}
\caption{\small $P(0,T)$ as a function of $T$. \label{fig:pricevsT}}
\makeatother
\includegraphics[width=3.6in]{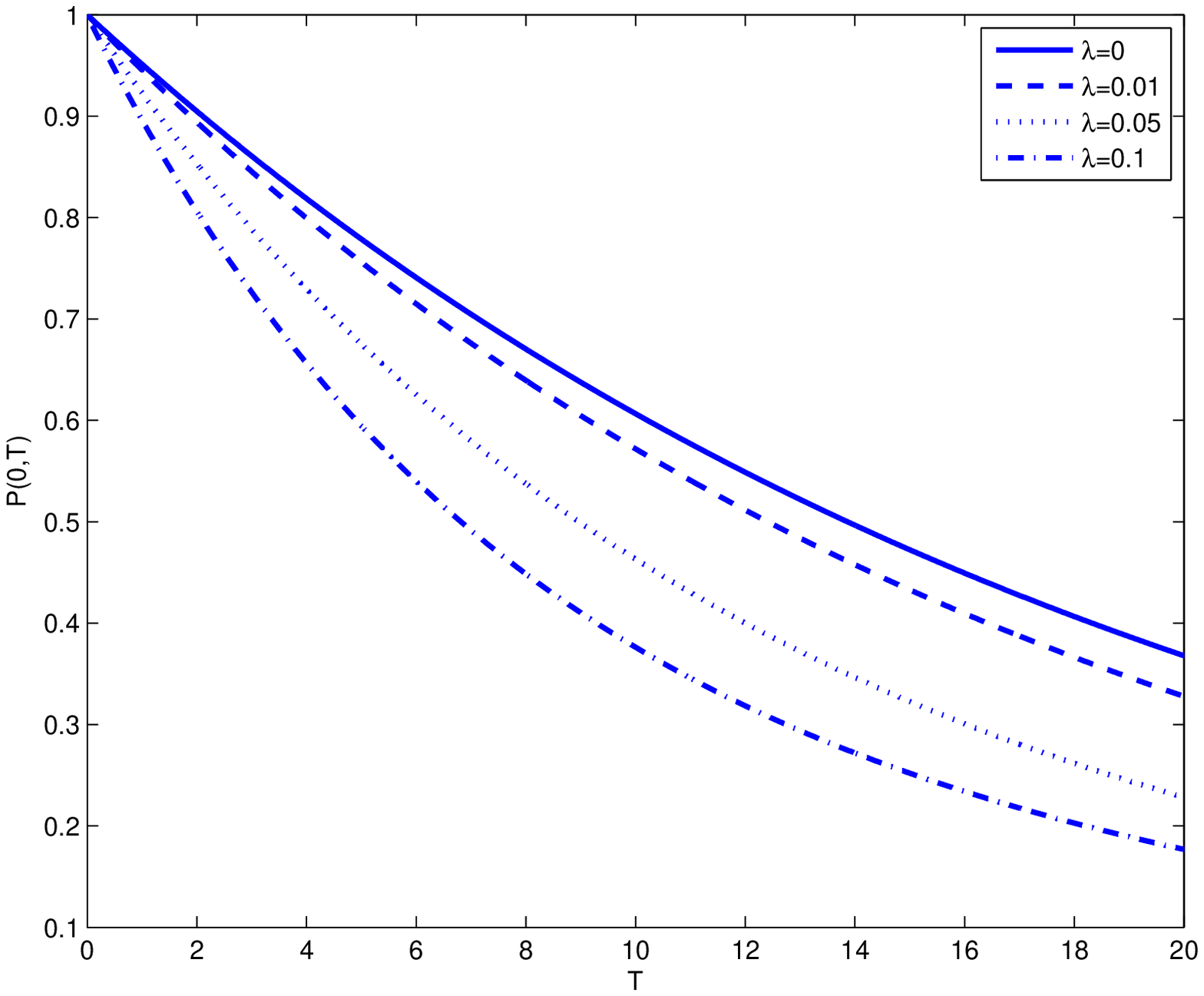}

\end{center}


\end{document}